\documentclass[sigconf]{acmart}

\usepackage{multirow}

\DeclareGraphicsExtensions{.ai,.pdf,.jpg,.png}
\setkeys{Gin}{pagebox=artbox}% Use artbox instead mediabox default. Possible values: mediabox, cropbox, bleedbox, trimbox, artbox. (shell: pdfinfo -box <pdf-file>)
\graphicspath{{.}}

\copyrightyear{2026}
\acmYear{2026}
\setcopyright{cc}
\setcctype{by}
\acmConference[CHIIR '26]{2026 ACM SIGIR Conference on Human Information Interaction and Retrieval}{March 22--26, 2026}{Seattle, WA, USA}
\acmBooktitle{2026 ACM SIGIR Conference on Human Information Interaction and Retrieval (CHIIR '26), March 22--26, 2026, Seattle, WA, USA}

\acmDOI{10.1145/3786304.3787868}
\acmISBN{979-8-4007-2414-5/2026/03}


\begin{document}

\title{Does Cognitive Load Affect Human Accuracy in Detecting Voice-Based Deepfakes?}

\author{Marcel Gohsen}
\email{marcel.gohsen@uni-weimar.de}
\orcid{0000-0002-1020-6745}
\affiliation{%
  \institution{Bauhaus-Universität Weimar}
  \city{Weimar}
  \country{Germany}
}

\author{Nicola Libera}
\email{nicola.lea.libera@uni-weimar.de}
\orcid{0009-0005-8775-3914}
\affiliation{%
  \institution{Bauhaus-Universität Weimar}
  \city{Weimar}
  \country{Germany}
}

\author{Johannes Kiesel}
\email{johannes.kiesel@gesis.org}
\orcid{0000-0002-1617-6508}
\affiliation{%
  \institution{GESIS - Leibniz Institute for the Social Sciences}
  \city{Cologne}
  \country{Germany}
}

\author{Jan Ehlers}
\email{jan.ehlers@uni-weimar.de}
\orcid{0000-0002-4475-2349}
\affiliation{%
  \institution{Bauhaus-Universität Weimar}
  \city{Weimar}
  \country{Germany}
}

\author{Benno Stein}
\email{benno.stein@uni-weimar.de}
\orcid{0000-0001-9033-2217}
\affiliation{%
  \institution{Bauhaus-Universität Weimar}
  \city{Weimar}
  \country{Germany}
}

\renewcommand{\shortauthors}{Gohsen et al.}

\begin{abstract}
Deepfake technologies are powerful tools that can be misused for malicious purposes such as spreading disinformation on social media. The effectiveness of such malicious applications depends on the ability of deepfakes to deceive their audience. Therefore, researchers have investigated human abilities to detect deepfakes in various studies. However, most of these studies were conducted with participants who focused exclusively on the detection task; hence the studies may not provide a complete picture of human abilities to detect deepfakes under realistic conditions: Social media users are exposed to cognitive load on the platform, which can impair their detection abilities. In this paper, we investigate the influence of cognitive load on human detection abilities of voice-based deepfakes in an empirical study with 30~participants. Our results suggest that low cognitive load does not generally impair detection abilities, and that the simultaneous exposure to a secondary stimulus can actually benefit people in the detection task.
\end{abstract} 

\maketitle

\section{Introduction}

Deepfakes refer to synthetic media content that is the result of manipulating source media in order to depict situations that did not occur in reality. Among the most popular applications of deepfake technology is to ``fake'' people by replicating their appearance or their voice \cite{mirsky:2022}. While there are beneficial applications, such as depicting historical figures \cite{wynn:2021} or objects \cite{lopez-garcia:2024, zaramella:2023} in a museum, deepfakes are predominantly used unethically and without consent of the depicted individuals \cite{deeptrace:2019}. Examples of malicious deepfake attacks include fraud \cite{bateman:2020, de-rancourt-raymond:2022}, non-consensual depiction in pornographic contexts \cite{wang:2022a, mania:2024}, and the spread of disinformation \cite{al-khazraji:2023}.  

One way to reduce these negative consequences of the malicious use of deepfakes is to educate individuals on how to recognize manipulated media \cite{shu:2020}. However, with the advancement of deep learning technologies, these deepfakes are getting hard to spot. Various studies on the human ability in detecting voice \cite{muller:2022, barrington:2025, han:2024, mai:2023}, video \cite{josephs:2023, groh:2022, prasad:2022, somoray:2023}, text \cite{chong:2023}, and multimodal deepfakes \cite{cooke:2024, frank:2024, groh:2024, hashmi:2024} come to similar conclusions: human detection performance of deepfakes is often not much better than a coin toss. 

Social media platforms are especially popular channels for spreading disinformation and fake news (e.g., through the use of deepfakes) \cite{lazer:2018, zhang:2020a}. The consumption of social media causes cognitive load for users \cite{pittman:2023} in the form of divided attention across multiple posts and advertisements \cite{hodas:2012} or even information overload \cite{gomez-rodriguez:2014}. We believe that this cognitive load has the potential to interfere with the human detection performance of deepfakes.

To our knowledge, this paper represents the first study on the impact of cognitive load on the human detection abilities of voice-based deepfakes. We specifically investigate voice modality as it can be effectively used to influence people \cite{gohsen:2023}, was among the most difficult modalities to detect in previous studies \cite{groh:2024, cooke:2024}, and requires only a small amount of source material to create realistic-sounding clones of individuals. Although video is the primary modality in social media, voice clones pose a real threat in terms of spreading disinformation, as attackers can imitate news videos by adding cloned voices of newsreaders to widespread B-roll (``symbolic'') footage, which provides no visual clues to spot a manipulation. 

This paper reports on a study involving 30~participants in which their ability to distinguish cloned voices of newsreaders from their real counterparts is examined in two experiments. In Experiment~1, an audio stimulus is presented to a participant, who has to decide whether this stimulus is real or fake under varying cognitive loads. We emulate this cognitive load with a 1-back task that has to be solved while listening to the stimulus. In Experiment~2, we create a different kind of distraction by showing participants related B-roll footage in parallel to the audio stimulus.%
\footnote{Stimuli are available at: \href{https://webis.de/data/webis-newsreader-voice-clones-26.html}{webis.de/data/webis-newsreader-voice-clones-26.html}} 
We find that light cognitive load, as caused by a 1-back task, does not in general impact human accuracy in detecting voice-based deepfakes. Furthermore, we observe that the simultaneous consumption of a secondary stimulus, such as the B-roll videos in Experiment~2, causes participants to get significantly better at the detection task.

In this paper, we contribute an extensive analysis of the body of literature that conducted empirical studies to examine human voice-clone detection abilities. Further, we outline a realistic attack that makes use of voice clones to spread disinformation on social media. Then, we describe a pipeline to obtain realistic-sounding voices-clones of real newsreaders and present our study. Finally, we report and discuss our findings and possible explanations.

\section{Related Work}
\label{related-work}

In the literature, we found 12 papers that report about studies that investigate and quantify the human ability in distinguishing between real (also known as {\it bona fide}) or fake (also known as {\it spoofed}) audio stimuli. A recent overview about those studies was provided by \citet{amirkhani:2025a}. In the following, we discuss and compare these studies with respect to experimental design, stimuli design, variables, findings, and identified indicators. The indicators are organized into a novel taxonomy.

\paragraph{Experimental Approaches}
An established procedure is to ask participants to listen to a single audio stimulus and assign one of two labels to the audio clip: bona fide or spoofed \cite{muller:2022,watson:2021,mai:2023,han:2024,barrington:2025,groh:2024,sharevski:2024,warren:2024}. For their study, \citet{alali:2025} mixes bona fide and spoofed audio into a single clip and hence require a third assignable class (``partially fake''). Some studies take into account the confidence of participants in their decision-making. \cite{bhalli:2024} enable participants to choose the option ``unsure'' instead of making a decision. \citet{barnekow:2021} allow participants to assign one of five labels, namely ``real'', ``rather real'', ``rather fake'', ``fake'' and ``no idea''. Similarly, \citet{frank:2024} ask their participants to rate deepfaked media (including speech) on a 5-point Likert scale ranging from ``definitely non-human'' to ``definitely human.'' 
In addition to the standard binary classification, \citet{barrington:2025} presents participants with two audio stimuli and ask them to decide if they originated from the same identity. These pairs of audio stimuli originate from either two different human speakers, two identical human speakers, or one human speaker and their cloned voice. A similar pairwise task is included in the study procedure of \citet{mai:2023}, in which participants are shown a pair of bona fide and spoofed audio stimuli and asked to identify which is which. 

A notable distinction between these studies is the format in which they are conducted. Eight of these studies are conducted as an online survey (e.g., crowdsourced via Prolific%
\footnote{\url{www.prolific.com}%
}) \cite{alali:2025,barrington:2025,barnekow:2021,frank:2024,groh:2024,mai:2023,muller:2022,warren:2024}. Although this format allows researchers to acquire numerous participants---between 102 \cite{barnekow:2021} and 3,002 participants \cite{frank:2024}---the acoustic properties of the environment (e.g., speaker vs. headphones, background noise) can not be sufficiently controlled. Two studies collect their data via web conferencing services (e.g., Zoom) \cite{han:2024, sharevski:2024}. While \citeauthor{han:2024} let participants play the audio stimuli on their local computer through the browser, \citeauthor{sharevski:2024} share the audio through the web conferencing software, which could introduce compression artifacts that might interfere with the experiment. Lastly, two studies are conducted in a controlled laboratory environment \cite{bhalli:2024,watson:2021}.

\paragraph{Experimental Stimuli}

When conducting a study on the human ability to detect auditory deepfakes, it is common practice to use a bona fide and spoofed stimuli from the same speaker. Previous studies have employed one of three methodologies to obtain such stimuli. The first method uses existing datasets that contain such stimuli. The second method involves using an existing dataset of bona fide stimuli, the speakers of which are spoofed using a state-of-the-art voice clone model. The third method creates a corpus of bona fide and spoofed stimuli, which may entail recording or crawling custom audio data. 

Popular datasets for these studies, which contain both bona fide and spoofed stimuli, originate from the Automatic Speaker Verification Challenge%
\footnote{\url{https://www.asvspoof.org/}}
(ASVspoof). Specifically, the datasets of the ASVspoof challenge of the years 2017 \cite{kinnunen:2017}, 2019 \cite{wang:2020}, and 2021 \cite{liu:2023a} have been used as target stimuli in the studies consulted \cite{bhalli:2024,han:2024,muller:2022, sharevski:2024,warren:2024}. These datasets represent a substantial database of speech recordings from the VCTK dataset \cite{yamagishi:2019} and spoofed audios, generated with a broad range of state-of-the-art TTS, voice conversion, and deepfake models. \citet{alali:2025} and \citet{barrington:2025} resort to their own previously created datasets from 2024, namely the RFP \cite{alali:2024} and the Deepspeak dataset \cite{barrington:2024}, respectively. The RFP dataset is a collection of bona fide audios obtained from the VCTK, the UK and Ireland English speech \cite{demirsahin:2020}, the DiPCo \cite{segbroeck:2020}, and the YouTube-8M \cite{abu-el-haija:2016} datasets which were used as target speakers for TTS, voice conversion, and partial deepfake approaches to produce spoofed audios. The Deepspeak dataset was created by asking crowdworkers to record a video of themselves speaking utterances, and then used audio and video manipulation such as voice cloning and lip-sync to generate spoofed media. \citet{bhalli:2024} use the ``Fake or Real'' (FoR) dataset \cite{reimao:2019} from 2019 in their study, which comprises bona fide stimuli from the CMU ARCTIC \cite{kominek:2003}, LJ Speech \cite{ito:2017}, and Voxforge \cite{maclean:2018} datasets and synthetic speech from commercial TTS systems. Since \citet{groh:2024} mostly focus on video deepfakes in the political domain, they use the Presidential Deepfakes dataset \cite{sankaranarayanan:2021} from 2021 to obtain stimuli for their study. Next to the ASVspoof 2021 dataset, \citet{warren:2024} also use the Wavefake \cite{frank:2021} and FakeAVCeleb \cite{khalid:2021} datasets (both created in 2021) in their study.

\paragraph{Factors Influencing the Recognition Performance}

The aforementioned studies examined the impact of different variables on the accuracy with which humans recognize voice clones. The studies assume that four main factors could influence human accuracy: the characteristics of the participant, the speaker, the spoken content, and the synthesis.

In terms of participant characteristics, the impact of demographics such as gender \cite{barrington:2025, bhalli:2024}, age \cite{barrington:2025, frank:2024, muller:2022}, educational background  \cite{frank:2024,watson:2021}, and country of residence \cite{frank:2024} have been analyzed. Additionally, the effect of language proficiency characteristics such as native language \cite{bhalli:2024,muller:2022} and level of fluency in a second language \cite{bhalli:2024} have been investigated. Furthermore, the levels of computing experience of a participant has been considered in three studies \cite{bhalli:2024,muller:2022,watson:2021}. Two studies focus on the human detection abilities of blind and visually impaired individuals \cite{han:2024, sharevski:2024}. Other studies investigate the impact of familiarity with the speaker's voice \cite{alali:2025,mai:2023} and of the level of training \cite{bhalli:2024,mai:2023,muller:2022}. 

The speaker's characteristics are for the most part underexplored in the aforementioned studies. So far, only the effects of a speaker's gender and their spoken dialect have been examined \cite{alali:2025}.  

An investigation of the impact of spoken content characteristics has been conducted by \citet{watson:2021}. In their study, they analyze whether the level of complexity (e.g., tongue twisters versus simple sentences) of voice clones has an impact on human detection abilities. Furthermore, they test whether mentioning a political candidate in the speech affects human accuracies. Finally, two studies examine the effects of spoken language on the human detection accuracy, in order to ascertain whether the language of the spoken content has an effect on this. While \citet{mai:2023} draw parallels between English and Mandarin, \citet{frank:2024} compare English, German, and Mandarin.

The last category of variables that are varied in studies of human voice clone detection abilities are characteristics of the synthesis. These characteristics include the choice of voice synthesis system \cite{alali:2025, sharevski:2024, warren:2024}, the type of the synthesis (e.g., TTS versus deepfake) \cite{han:2024,muller:2022}, and duration of the synthesized audio \cite{barrington:2025,mai:2023,watson:2021}. Of the analyzed studies, \citet{frank:2024} and \citet{groh:2024} also consider different kind of modalities of the synthesized deepfakes such as video and text.

\paragraph{Study Findings}

As the human performance in detecting voice clones depends on a variety of factors, the measured human accuracy varies dramatically from study to study. A large proportion of the studies find that human accuracy in detecting voice clones range between 60\% and 80\% \cite{barrington:2025,groh:2024,mai:2023,muller:2022,warren:2024}. Below that range, \citet{frank:2024} find accuracies slightly above chance of between 50\% and 60\% depending on a participant's nationality and the language of the audio stimuli. Accuracy is measured in a comparable range by \citet{han:2024} and \citet{sharevski:2024}---both studies that investigate the detection performance of blind or low vision individuals. However, there are outliers reporting on accuracies that are considerably below chance. The study by \citet{watson:2021} reports on an average accuracy of 42\% across university students. Similarly, \citet{barnekow:2021} clone the voice of a university professor familiar to their participants and find that they could only recognize the cloned voice correctly in 37\% of the cases. The lowest found accuracy is 16\% for detecting partial fake speech (i.e., single media items that contain both bona fide and spoofed voices) \cite{alali:2025}. 

In addition to detection accuracies, studies investigated which factors significantly impact these values. One of the most detrimental factor is what approach was used to create the spoofed stimuli \cite{warren:2024} which can cause a difference of up to 20\% in average accuracy. Furthermore, \citet{muller:2022} find that native speakers have an advantage in detecting voice clones speaking in their native language. According to \citet{watson:2021}, the spoken content has a major impact; complex spoken content is easier to recognize than simple content in spoofed audios. Finally, \citet{bhalli:2024} find that training sessions can help to increase the human detection performance.

On the contrary, some studies also identified factors that do not affect the human detection performance. \citet{barrington:2025} do not find any effect of the speaker's gender. On the participant side, minor effects are associated with demographics \cite{frank:2024} except the performance decline with increased age \cite{muller:2022}. Particularly, an educational background in computing does not affect the human detection performance \cite{muller:2022,watson:2021,bhalli:2024}. Lastly, the listening behavior of a participant does not play a significant role such as the number of stimulus repetitions or the time spend on the detection task \cite{mai:2023}.

\paragraph{Indicative Characteristics of Stimuli}

\begin{figure}[t]
\small
\setlength{\itemsep}{0.5ex}
\begin{minipage}[t]{.48\linewidth}
\vspace*{-0pt}
\begin{enumerate}
\item Artifacts
\begin{itemize}
\item Background noises \cite{barrington:2025,barnekow:2021}
\item Frequency profile \cite{barnekow:2021}
\item Recording quality \cite{barrington:2025,han:2024}
\item Reverb and echo \cite{barnekow:2021,han:2024}
\end{itemize}

\item Inflection
\begin{itemize}
\item Emotions \cite{barrington:2025,han:2024}
\item Intonation \cite{mai:2023}
\item Monotonousness \cite{barrington:2025}
\item Roboticness \cite{mai:2023}
\end{itemize}
\end{enumerate}
\end{minipage}%
\hfill
\begin{minipage}[t]{.48\linewidth}
\vspace*{-0pt}
\begin{enumerate}
\setcounter{enumi}{2}
\item Pronunciation
\begin{itemize}
\item Accents \cite{barrington:2025,han:2024}
\item Filler words \cite{han:2024}
\item Mispronunciations \cite{barrington:2025,han:2024}
\item Stutters \cite{barrington:2025}
\end{itemize}

\item Rhythm
\begin{itemize}
\item Breathing \cite{barrington:2025,han:2024}
\item Pace \cite{barrington:2025,mai:2023}
\item Pauses \cite{barrington:2025,han:2024,mai:2023} 
\end{itemize}       
\end{enumerate}
\end{minipage}
\caption{Our taxonomy of indicative characteristics of stimuli that participants in various studies reported using to distinguish between bona fide and spoofed voices. The categories are detailed in the last part of Section~\ref{related-work}}
\label{taxonomy-of-indicators}
\Description{This figure shows a taxonomy of indicative characteristics of stimuli that participants in various studies reported using to distinguish between bona fide and spoofed voices. The indicators are divided into four main groups: (1) artifacts, (2) inflection, (3) pronunciation, and (4) rhythm.}
\end{figure}

In the different studies that we analyze, participants report on different indicators that helped them to distinguish between bona fide and spoofed stimuli. In Figure~\ref{taxonomy-of-indicators}, we organize these properties into a taxonomy to provide an overview of these indicators. 

The artifacts group (1) comprises acoustic artifacts that are unwanted byproducts of speech synthesis algorithms, which include background noises, unnatural frequency profiles, varying levels of perceived audio quality, and inconsistent reverb and echo behavior. Most of these lead participants to decide that a stimulus is spoofed. 

The inflection group (2) is concerned with the ``melody'' and stress of the speech. The existence or lack thereof of perceived emotions can be an indicator of either bona fide or spoofed audio stimuli, respectively. Intonation can be a tool to express such emotions by varying the pitch of pronounced syllables in an utterance. If these intonations are underexpressed or unnaturally frequent, participants perceived stimuli as monotone or robotic.

The pronunciation group (3) describes how the words in an utterance are pronounced. Accents (e.g., New York accents) influence this pronunciation and can impact whether a participant perceive an audio as bona fide or spoofed. Added filler words and stutters are usually associated with bona fide stimuli, while odd mispronunciations can be indicators for spoofed stimuli as well.

Finally, the rhythm group (4) describes temporal properties of a spoken utterance. Breathing at anticipated times, although reproducible with modern voice synthesis algorithms, are typically indicators for bona fide stimuli. The same is true for pauses in the speech. The overall pace of a spoken utterance can also give away if a stimulus is spoofed or bona fide.

\section{Fake News Attack Model}
\label{fake-news-attack}

To demonstrate how voice-based deepfakes can be instrumentalized to spread disinformation, we outline a real-world attack model. This attack model is based on the premise that newsreaders, journalists and social media influencers publish numerous high-quality videos in which they report on news on social media platforms or news portals. These videos are easily accessible to potential attackers, who can use voice cloning to imitate the voices of these speakers once they have collected enough high-quality material. For example, Elevenlabs%
\footnote{\url{https://elevenlabs.io}} 
suggests a minimum amount of source material of 30~minutes.
The attackers could use the cloned newsreader voices to report fake news that fits their agenda, which would appear believable due to the credibility of the imitated newsreader. When news is presented in a voice-over style, combined with symbolic B-roll video footage, with the newsreader not visible, all the visual cues indicating that the media is fake are eliminated. 

\begin{figure}[t]
\includegraphics[width=\linewidth]{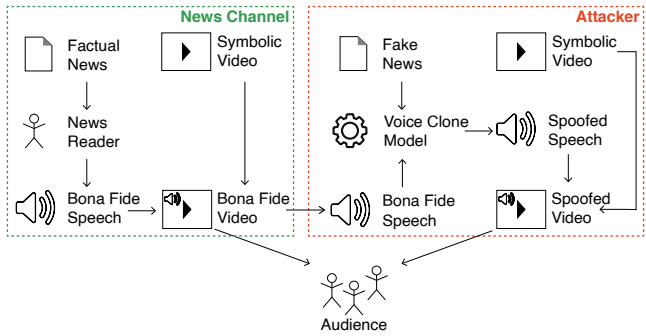}
\caption{Real-world attack model for spreading fake news videos with voice-based deepfakes. The audience is tricked into believing to hear some well-known newsreader---and thus that the news is spread by the reader's news outlet. Our study is designed to mirror corresponding attacks.}
\Description[Real-world attack model for spreading disinformation]{The figure illustrates the roles of a news channel and an attacker in the process of disinforming an audience. The news channel turns a piece of factual news item into bona fide speech through a newsreader and pairs it with symbolic video material to create a bona fide news video which is made public. The attacker extracts the speech from the public video, feeds it into a voice clone model, and synthesizes spoofed speech together with a piece of fake news. The spoofed speech combined with symbolic video footage represents a spoofed news video that is released into the public to disinform an audience.}
\label{voice-clone-fake-news-attack}
\end{figure}

Figure~\ref{voice-clone-fake-news-attack} outlines the attack model differentiating the roles of a news channel, an attacker, and an audience. On the side of the news channel, a newsreader reads out factual news in voice-over style, which is combined with symbolic video material to create the bona fide news video. The attacker collects audio material through, for example, the extraction of audio channels from news videos and other available sources and then uses it to train or condition a voice clone model. This model generates a voice-over narration for attacker-provided fake news in the newsreader's voice, which is paired with original news footage or a symbolic video to produce the spoofed news video ready for publication on social media. 

This attack model is not just a thought experiment, but is currently being used in real attacks. In the beginning of 2025, journalist Georgina Findlay became victim of such an attack, in which far right social media channels cloned her voice to spread disinformation following fascist ideologies \cite{findlay:2025}. In 2023, German news program ``Tagesschau'' was under attack, in which the voice's of their newsreaders were cloned to spread disinformation about the Ukrainian war and the Covid-19 pandemic \cite{bastian:2023}. 

\section{Stimulus Design}
\label{audio-stimuli}

To obtain bona fide and spoofed stimuli for our experiments, we follow a methodology that is plausible for potential attackers in the described attack model in Section~\ref{fake-news-attack}. 
We suspect that a potential attacker is likely to acquire source media from social media platforms. Therefore, we apply the same strategy, manually collecting videos from YouTube. The news videos that we deem usable are clips from well-known news channels (e.g., NBC, BBC) that are of high audiovisual quality and have low levels of background noise. As the effectiveness of the attack depends on the acoustic properties of a voice, we include different speakers in our experiments. Specifically, we collect videos of four renowned newsreaders (two female and two male). The following four newsreaders are selected based on their vocal diversity and the amount of obtainable high-quality source media on YouTube:

\begin{itemize}
\item Andrea Mitchell. {\it American journalist at NBC News}.
\item Carl Nasman. {\it American journalist at BBC News}.
\item Lester Holt. {\it American journalist at Dateline NBC}.
\item Sophie Raworth. {\it English journalist at BBC News}.
\end{itemize}

The source videos undergo an extensive manual filtering and editing process. Duplicate video segments are removed, and noisy parts are cut out. This collection, filtering, and editing process is repeated until at least one hour's worth of training data has been collected for each of the four voices. During this process, suitable candidate segments for bona fide stimuli in the study are selected and set aside, i.e., not used for training the voice clone model. 

We apply additional automatic preprocessing steps to the training portion of the videos. Music and background noises are removed by employing an MDX-Net model \cite{kim:2021a} for music demixing and a model called ``Kim Vocal 1'' from the Ultimate Vocal Remover v55%
\footnote{\url{https://ultimatevocalremover.com/}}
application for the isolation of the voice. The result of this step is a clean vocal track of the video. To make the training process more efficient, these audio clips are split into semantically coherent segments of less than 10~seconds using an automatic audio splitter that uses WhisperX \cite{bain:2023}, a long form audio transcription approach, to find appropriate segmentations.%
\footnote{\url{https://github.com/JarodMica/audiosplitter_whisper}}
Following this step, we obtained 2,872 audio clips, each averaging about six seconds in lengths. 

In initial experiments with various open-source Text-to-Speech systems (TTS) that feature voice cloning, such as F5-TTS \cite{chen:2025}, StyleTTS~2 \cite{li:2023a}, TorToise \cite{betker:2023}, and VoiceCraft \cite{peng:2024}, TorToise produced the most authentic reproduction of the four selected voices. TorToise is an autoregressive transformer combined with a diffusion-based denoising algorithm and a MEL-based vocoder. Another advantage of TorToise is that its GUI-based training process%
\footnote{\url{https://github.com/JarodMica/ai-voice-cloning}}
is sufficiently simple, so that potential attackers do not require expert skills in AI~applications to clone a voice. 

We train a model for each of the four selected speakers, with the help of the aforementioned GUI. Each model is trained with a batch size of~80 for 400~epochs with model checkpointing for every~50 epochs. We then use each of the resulting models to generate example outputs and to manually determine the ideal number of training epochs per voice. For the voices of Lester Holt and Sophie Raworth, 300 epochs yielded the best results, whereas for Andrea Mitchell and Carl Nasman, 350 training epochs were preferable. In terms of training hyperparameters, the default values as suggested by the training GUI are mostly used, such as a learning rate of~$10^{-5}$. However, some optimizations of the training hyperparameters were performed in the initial experiments, resulting in MEL and text learning rate weights of~$1.0$ and~$0.6$, respectively. 

In order to eliminate the influence of spoken content and limit human detection characteristics to acoustic features, the bona fide stimuli are transcribed, and the transcriptions are used to reproduce the same content with the trained voice clone models. The generated spoofed stimuli then undergo further manual post-processing. All artifacts and noise are filtered out, and the frequency response curves are brought closer to the original audio signal by applying equalizer curves. Finally, the volume of the bona fide and spoofed stimuli is normalized to ensure identical loudness levels. 

In total, we prepare 12~audio stimuli for each voice, six of which are bona fide and six of which are spoofed (48~in total). The audio clips are between 11~and 15~seconds long and correspond to two to three sentences of spoken content. All audio clips are stored as lossless PCM waveforms with a sampling rate of 44.1~kHz. 

For Experiment~2, in which videos are shown to participants while listening, the pool of audio stimuli is extended with additional stimuli following the methodology described above. For each of the four selected newsreaders, we select two news videos that are publicly available. To avoid providing any visual cues whether a stimulus contains bona fide or spoofed audio, we collect video material in which the speaker is not visible. The collected videos feature symbolic visualizations to engage the viewer (sometimes referred to as B-roll footage \cite{huber:2019}), while the news are provided as a voice-over.
In total, we create a pool of 16~video clips, 8~of which have unchanged audio tracks and 8~of which involve spoofed audio. The clips are between 12~and 15~seconds long. 
\section{Empirical Study}

\begin{figure*}[t]
\includegraphics[width=\linewidth]{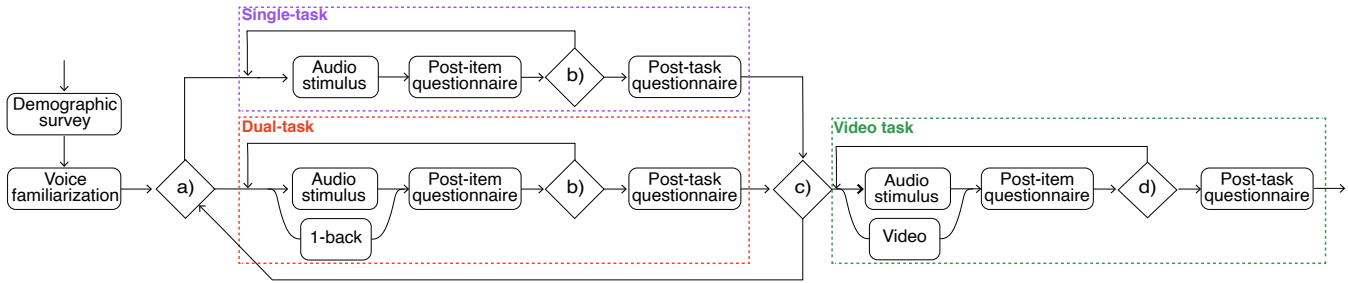}
\caption{Flow chart of the conducted study consisting of single-task and dual-task conditions of Experiment~1 and the video condition in Experiment~2. The decisions a) and c) are mechanisms to randomize the order of conditions and ensuring that both conditions have been performed once. The decision b) and d) repeat the listening task for 24 and 8 different stimuli in Experiment~1 and 2, respectively.}
\Description{This figure depicts a flowchart of the individual steps of the study in the two experiments. Participants complete a demographic survey and familiarize themselves with the voices before they perform Experiment 1 under the single-task or dual-task conditions. They repeat Experiment 1 under the missing condition and proceed to Experiment 2 which features the video task.}
\label{chiir26-study-flow-chart}
\end{figure*}

In order to investigate the extent to which humans are capable of recognizing the attacks mentioned above, we conducted two experiments. Experiment~1 is a laboratory-controlled study in which the cognitive load is systematically varied using a dual-task scenario. The aim is to examine how this affects the ability to distinguish between bona fide and spoofed voices. However, recent research shows that humans are less accurate at detecting deepfakes in audio-only scenarios compared to audiovisual media \cite{groh:2024,cooke:2024}. We therefore opted for a simple secondary task (similar to the established 1-back assignment \cite{grimmer:2021}) to ensure the primary discrimination task (bona fide versus spoofed voice) could still be carried out. Experiment~2 constitutes an application-oriented extension of the research design, with the objective of validating the results using real-life video sequences. The methodological approach employed in both experiments is outlined below.

\subsection{Experiment~1: Auditory Sequences}

Experiment~1 uses a dual-task technique to control and vary the utilization of working memory capacities (cognitive load). The aim is to investigate how this factor influences the reliability with which bona fide and spoofed voices can be distinguished.

\subsubsection{Design and Procedure}

Experiment~1 takes place in a small laboratory room. Participants are greeted and requested to sign a form of consent. The experimental procedure is depicted in Figure~\ref{chiir26-study-flow-chart} and described below. Participants are asked to complete a demographic survey through which they could specify their age and gender, and rate their experience with artificial voices and their confidence in detecting them. The latter two were collected on a 7-point Likert scale ranging from no expertise~(1) to expert~(7), and from not at all confident~(1) to very confident~(7), respectively. 

Following this, participants are asked to familiarize themselves with the bona fide voices of the four newsreaders. For each speaker, we provide an approximately 15-second audio clip. Participants are asked to listen to all four clips from beginning to end at least once. The respective clips are not included in the pool for the single- and dual-task exercises. Once participants become used to the voices, the experiment starts.

To vary the working memory load, participants perform a single and a dual task. The order of the assigned condition is counterbalanced across participants. In the single-task condition, participants are asked to listen to 24~audio clips that are played in random order via speakers. After listening to each clip, they are asked to indicate whether the stimulus is bona fide or spoofed and to describe what leads them to their decision. In addition, they are asked to rate their confidence on a 7-point Likert scale and to evaluate the perceived quality of the audio clips. At the end of each task, participants are asked to rate the perceived difficulty and report heuristics that helped them distinguish between bona fide and spoofed stimuli. In the dual-task condition, a secondary task has to be completed along with the primary task. Specifically, participants are required to observe a series of digits displayed on a monitor and have to press a button when a~``3'' follows a~``1.'' The numbers on the screen are updated at a rate of 0.8 seconds. The proportion of numbers that are target stimuli range from 13\% to 33\%.

The 48 audio stimuli (obtained as described in Section~\ref{audio-stimuli}) form a combined pool from which stimuli are randomly drawn for each participant in the single-task and dual-task conditions. However, we ensure that bona fide and spoofed audios of identical utterances never become stimuli for the same task. Participants are not informed about the distribution of bona fide and spoofed stimuli.

\subsection{Experiment~2: Video Sequences}

Experiment~2 is conducted with the same participants immediately after the completion of Experiment~1. However, while Experiment~1 uses an established approach to control and vary cognitive load, Experiment~2 incorporates videos as an application-oriented component, which constitutes the most common type of content on social media. Both experiments focus on the manipulation of auditory information. As in Experiment~1's dual-task condition, decision-making in Experiment~2 (bona fide vs.\ spoofed voices) is made more difficult by incorporating visual information, although no controlled variation of the cognitive load is performed here.

\subsubsection{Design and Procedure}

Experiment~2 follows a similar study procedure as Experiment~1. Participants look at a sequence of 8~videos, four of which containing spoofed audio, and have to indicate after being exposed to a stimulus whether the stimulus contains bona fide or spoofed audio. The post-item and post-task questionnaires in Experiment~2 are identical to questionnaires in Experiment~1. From the array of 16~video stimuli, 8~are selected at random, bearing in mind that videos reciting the same utterance (with bona fide and spoofed audio) are excluded from appearing within the same task for the same participant.

\subsection{Apparatus}

The study application is implemented as a Flask server%
\footnote{\url{https://flask.palletsprojects.com}}
responsible for data logging and assignment of stimuli and tasks. The front end is implemented using native HTML and JavaScript to provide survey forms and to collect data as well as to implement the stimulus-response architecture (Experiment~1) and media playback (Experiment~1 and~2). As the popularity of consuming social media content on mobile devices in speaker mode increases, participants listen to the stimuli through the internal dual speaker setup of an HP OMEN~17 laptop in both experiments.

\subsection{Participants}

30 volunteers (22 males, none of whom are non-binary, with an age between 20 and 36 years) participated in both experiments. Due to the composition of the sample, age-related limitations in hearing ability should be ruled out \cite{muller:2022}. Furthermore, none of our test subjects reported any issues with their ability to perceive sound. Written informed consent was obtained prior to the data collection.

\section{Results}

In the following, we discuss the results for the two experiments---Experiment~1 with auditory sequences and Experiment~2 with video sequences---with respect to the accuracy of the participants in detecting bona fide and spoofed voices.

\subsection{Results: Experiment~1}

\begin{figure}[t]
\centering
\includegraphics[width=.9\linewidth,trim={0 40pt 0 20pt}]{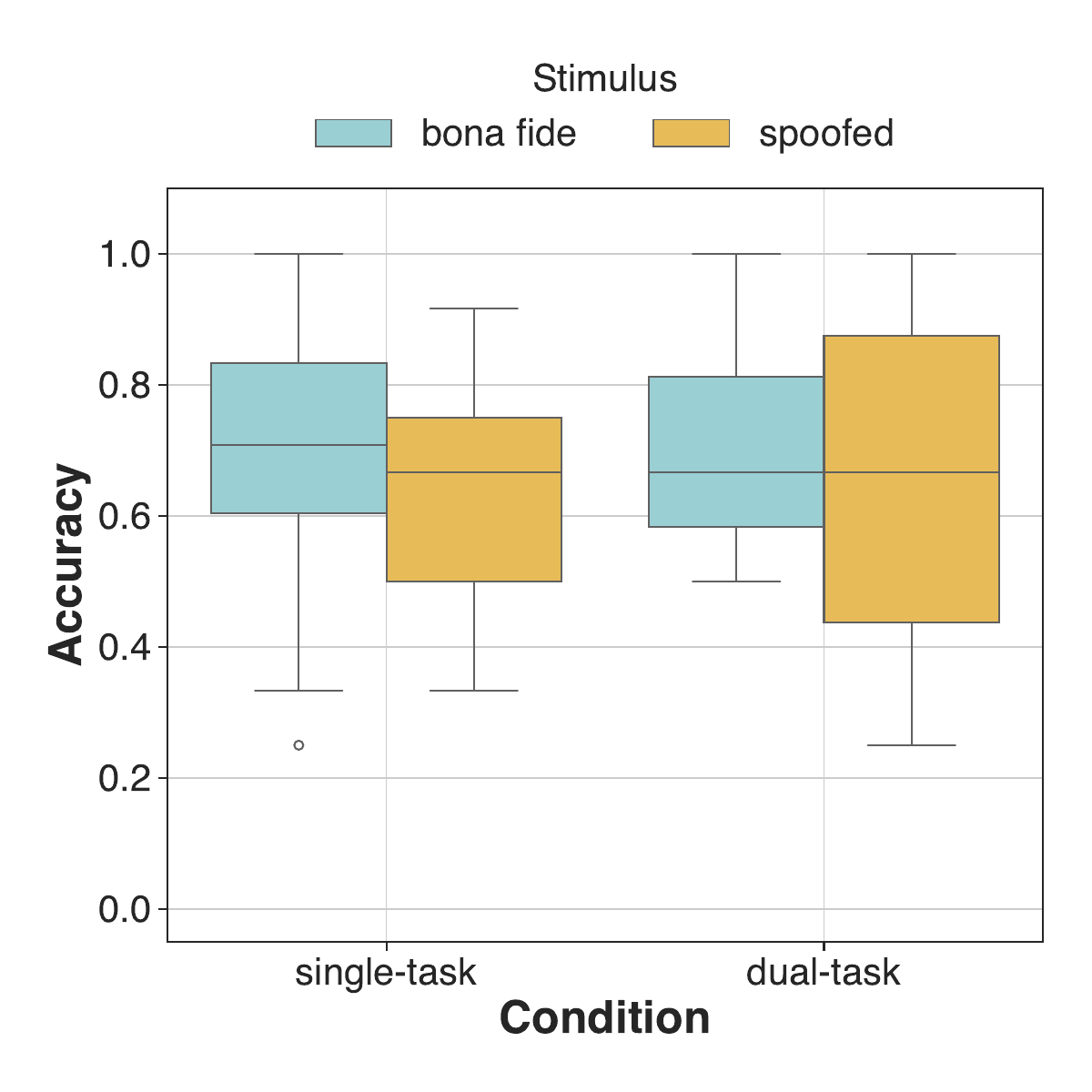}
\caption{Accuracy in detecting voice clones in the single- and dual-task conditions for spoofed and bona fide stimuli averaged by participant.}
\Description{The figure shows a box plot comparing accuracies in detecting voice clones between the single-task and dual-task conditions. Furthermore, the detection accuracies are grouped by bona fide and spoofed stimuli. The medians for the single-task conditions are about 0.7 for bona fide and 0.66 for spoofed stimuli. Under the dual-task condition, detection accuracy medians are 0.66 for both bona fide and spoofed stimuli.}
\label{accuracy-per-condition}
\end{figure}

Under single-task and dual-task conditions, participants achieve an average accuracy of 0.67. The detection accuracy is slightly lower under the dual-task ($\mu=0.66$, $\sigma=0.13$) than under the single-task condition ($\mu=0.68$, $\sigma=0.12$). According to a paired t-test, the mean accuracies achieved under these two conditions do no differ significantly ($p=0.6$). Independent of the task conditions, the mean accuracy in detecting bona fide stimuli is higher ($\mu=0.7$, $\sigma=0.14$) than for spoofed stimuli ($\mu=0.63$, $\sigma=0.18$), accompanied by a substantial increase in standard deviation. In Figure~\ref{accuracy-per-condition}, we compare the detection accuracy of bona fide and spoofed stimuli under individual experiment conditions. While the median accuracies are seemingly unaffected, we can see that the detection accuracies of spoofed stimuli are much less consistent between participants under the dual-task condition, which causes an increase of standard deviation from 0.17 to 0.24. On the contrary, participants are more consistent to detect bona fide stimuli under the dual-task condition, causing a drop in standard deviation from 0.19 to 0.14.

\begin{figure}[t]
\centering
\includegraphics[width=.9\linewidth,trim={0 40pt 0 20pt}]{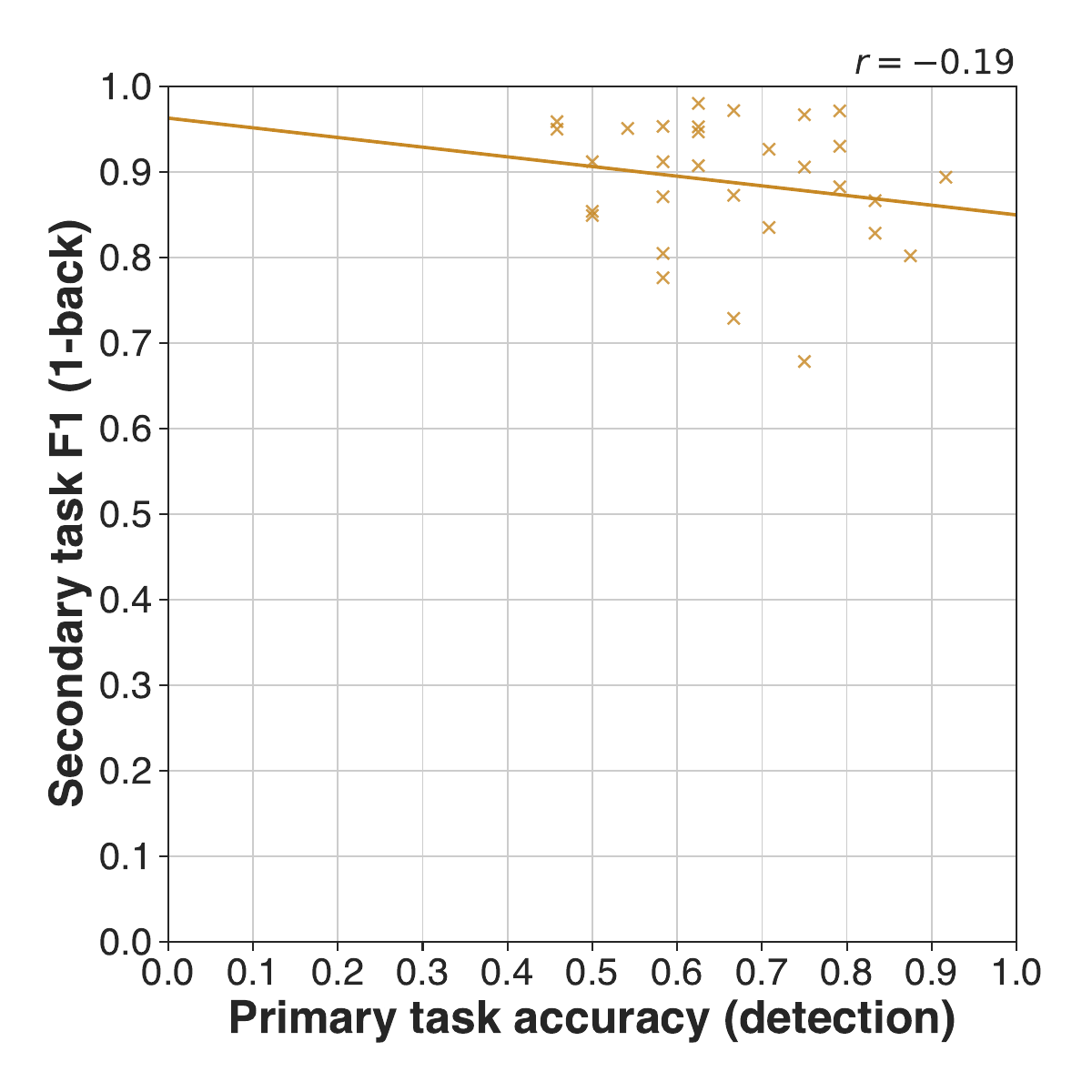}
\caption{Analysis of correlation between primary (voice-clone detection) task accuracy and secondary task (1-back task) $F_1$ of the participants. There is a weak negative correlation (Pearson $r=-0.19$) indicating that a trade-off of cognitive capacities exists.}
\Description{The figure shows a scatter plot with a regression line, with the accuracy of the primary task shown on the x-axis and the secondary task F1 value shown on the y-axis. The regression line shows a weak negative correlation.}
\label{dual-task-accuracy-correlation}
\end{figure}

As the mean accuracy seems to be unaffected by applying cognitive load, we examine if there is a trade-off in primary (detection) and secondary task (1-back) performances under the dual-task condition. If such a trade-off exists, participants exhibit a limited cognitive capacity, which is divided across the primary and secondary task. In Figure~\ref{dual-task-accuracy-correlation}, the participant's accuracy and $F_1$ in the individual tasks are compared.  According to Pearson's correlation coefficient, a weak negative correlation ($r=-0.19$) can be observed, which implies that participants trade off performance in the individual tasks. However, this correlation is not significantly different from zero ($p=0.32$).

\begin{table}[t]
\caption{Effects of various variables in predicting whether a participant guesses correctly according to a logistic regression model.}
\centering
\begin{tabular}{@{}l c c c@{}}
\toprule
\bf Variable & \bf Coefficient & \bf $z$-value & \bf $p$-value\\
\midrule
Decision confidence & \phantom{-}0.147 & \phantom{-}4.050 & \bf <0.001\\
Ground truth & \phantom{-}0.282 & \phantom{-}2.444 & \phantom{<}0.015\\
Speaker: Mitchell & -0.623 & -1.251 & \phantom{<}0.211\\
Participant gender & \phantom{-}0.143 &\phantom{-}1.079 & \phantom{<}0.281\\
Stimulus & -0.013 & -0.757 & \phantom{<}0.452\\
Participant age & \phantom{-}0.008 & \phantom{-}0.555 & \phantom{<}0.572\\
Condition & \phantom{-}0.052 & \phantom{-}0.454 & \phantom{<}0.650\\
Speaker: Nasman & -0.232 & -0.410 & \phantom{<}0.682\\
Speaker: Raworth & \phantom{-}0.185 & \phantom{-}0.220 & \phantom{<}0.826\\
Speaker: Holt & -0.028 & -0.041 & \phantom{<}0.967\\
Participant experience & -0.004 & -0.103 & \phantom{<}0.918\\
\bottomrule
\end{tabular}
\label{table-logisitic-regression}
\end{table}

To estimate the effects of individual variables, we fit a logistic regression model that predicts whether a participant made a correct decision. A Likelihood-ratio test on the model reveals that at least one of the variables significantly improves the model fit ($p=0.0003$). Table~\ref{table-logisitic-regression} contains the resulting coefficients, $z$-values, and $p$-values for each analyzed variable, where the $p$-values originate from a post hoc Wald-Test. To avoid significant findings due to chance caused by the problem of multiple comparisons, we perform an alpha correction using the Holm-Bonferroni method ($\alpha=0.005$ after correction). Decision confidence (i.e., how certain a participant is about their decision) has a significant influence, suggesting that participants have reflected well on their decisions. The ground truth (i.e., whether a stimulus is bona fide or spoofed) may or may not have a significant effect, depending on whether one accepts the conservative correction. The dummy variable for bona fide is~1, which means that the positive coefficient shows that if the stimulus is bona fide, there is a higher probability that the participant guesses correctly. The other variables have $p$-values that are undoubtedly consistent with the null hypothesis and we omit their interpretation.
%%% NOTES. (omitted)
%%% However, we can see some effects based on the cloned speaker with stimuli from Andrea Mitchell being the hardest to guess (mean accuracy $\mu=0.62$) and stimuli from Sophie Raworth being the easiest to get right (mean accuracy $\mu=0.70$). Participant demographics exhibit minor effects, where male participants were slightly more accurate ($\mu=0.67$) than female participants ($\mu=0.66$). Variables that show almost no systematic effect include the choice of stimulus (i.e., which exact stimulus out of 48 was shown), the condition (single-task versus dual-task) as well as a participant's self-rated experience with synthetic voices. 

\begin{figure}[t]
\centering
\includegraphics[width=.9\linewidth,trim={0 40pt 0 20pt}]{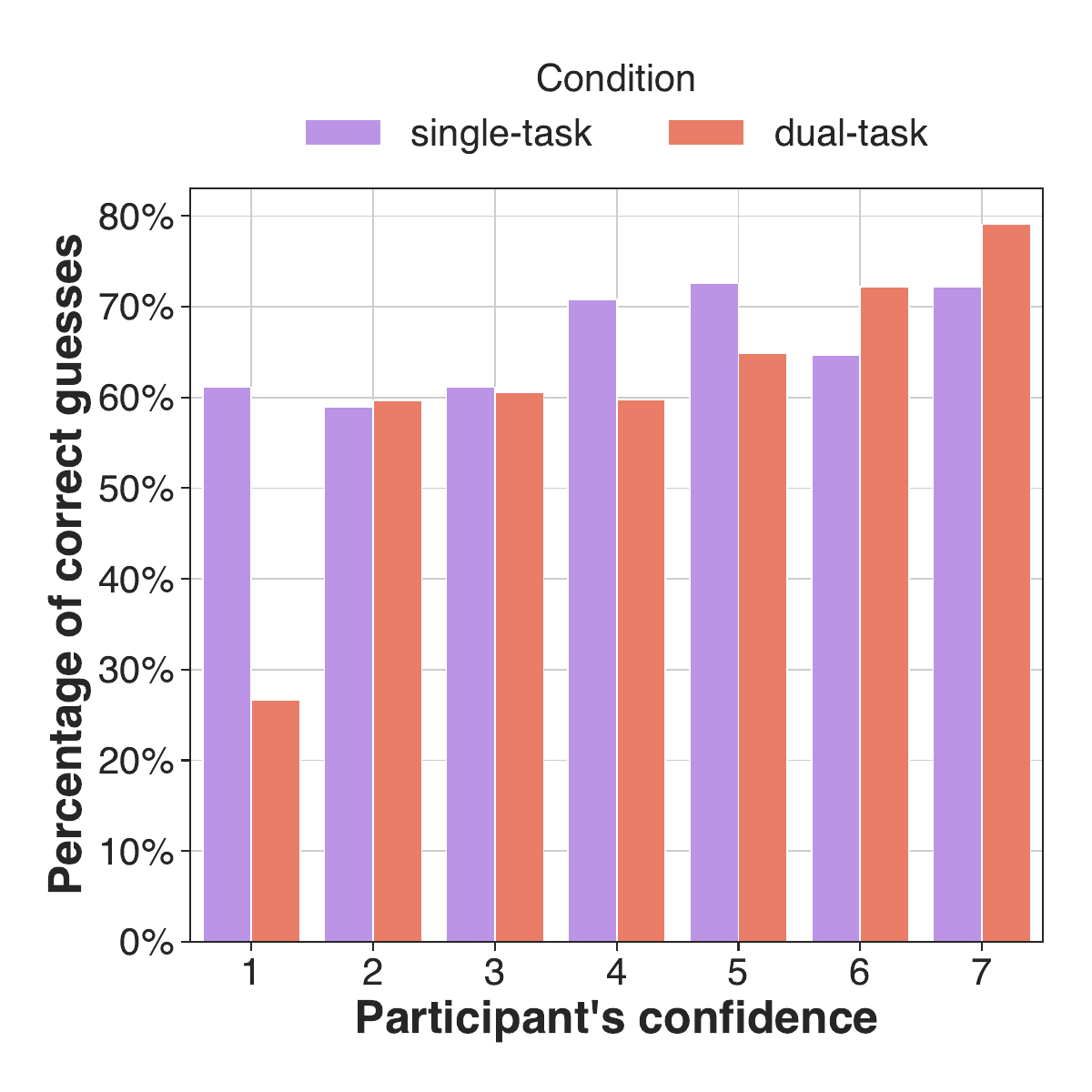}
\caption{Percentage of correct guesses by participants, broken down by their self-assessed confidence in their guess (higher score = greater confidence in the guess), separated by single-task and double-task conditions.}
\Description{The figure displays a bar plot with the self-rated participant confidence (7-point Likert scale) on the x-axis and the percentage of correct guesses on the y-axis. The bars are grouped by the single-task and dual-task conditions. Participants under the dual-task condition reflected better on their decision leading to only about 20\% correct decision at confidence value of one and almost 80\% of correct decisions at confidence value of seven. Under the single-task conditions the percentage of correct guesses are almost uniformly distributed across all seven confidence values.}
\label{confidence-correlation}
\end{figure}

As the results of the logistic regression identify a participant's confidence in the decision as a good predictor of the participant's success, we dig deeper to see how accurate the self-reflection is for the single- and dual-task conditions. Figure~\ref{confidence-correlation} shows the percentage of correct decisions under the condition that a participant assigned a specific confidence score. Under the dual-task condition, the function of the proportion of correct guesses resembles a monotonically increasing function based on a participant's confidence rating. However, a participant's self-reflection abilities under the single-task condition seem to be worse. A point-biserial correlation analysis reveals a correlation of $r=0.06$ and $r=0.17$ under the single- and dual-task conditions, respectively, where only the latter is significantly different from zero ($p\approx7\times 10^{-6}$). Participants under the single-task condition have a tendency to underestimate their detection abilities. Although they assigned the lowest possible confidence of one, they still guess correctly above chance (60\% correct guesses). The highest proportion of correct responses in the single-task condition are collected when participants assigned a confidence score of five.

\paragraph{Decision Indicators}

As part of the experiment, we asked participants for a rationale on every decision to find common indicators that humans use to decide between bona fide and spoofed. These free text answers are manually analyzed to obtain an overview. 

We find comparable indicators to what prior studies reported (see Figure~\ref{taxonomy-of-indicators}). Participants identified artifacts such as ``glitches'', drops in volumes, existence of digital effects (e.g., chorus, reverb, echo), static and interference which make participants vote for spoofed. However, also artifacts of human recording are identified such as smacking of lips, breathing or microphone pops that make participants believe that a stimulus is bona fide. Similar aspects are identified with respect to the frequency profile of a stimulus where dull, flat, muffled or distorted profiles are associated with spoofed and clear and scratchy voices with bona fide stimuli. Voices labeled as husk or hoarse are identified as either bona fide or spoofed. In terms of inflections, participants identify absence of emotions, robotic intonation, and monotone utterances as traits of spoofed stimuli. One participant stated that the utterance exhibited a relaxed way of speaking and decided for bona fide while another participant declared a voice as sounding irritated and labeled the stimulus as spoofed. Most reported indicators are about pronunciation aspects of the utterances, which in most cases is an indicator for spoofed stimuli. Participants categorize accents as fake and point out mispronunciations, mumbling, stumbling over words, and badly articulated syllables and word endings. Rhythm aspects are also present, where inconsistent pace, or too fast or too slow pace is associated with spoofed audio. Pauses are more frequently associated with bona fide stimuli.

Some indicators are found in our study that are not part of our taxonomy. One participant established a connection to the spoken content, even though identical utterances are used for bona fide and spoofed stimuli. Participants said that the content sounded like it was generated by AI or that the news content was implausible, which made them choose spoofed. This indicator aligns with an observation made by \citet{watson:2021} that if a political candidate appeared in the utterance, more participants decided it was fake. Furthermore, the familiarization seems to have an effect causing participants to justify their choices by saying that the stimulus sounded different (spoofed) or similar (bona fide) to the remembered voice. Finally, some decisions are justified by simply having a ``gut feeling'' or intuition.

\subsection{Results: Experiment~2}

\begin{figure}[t]
\centering
\includegraphics[width=.9\linewidth,trim={0 40pt 0 20pt}]{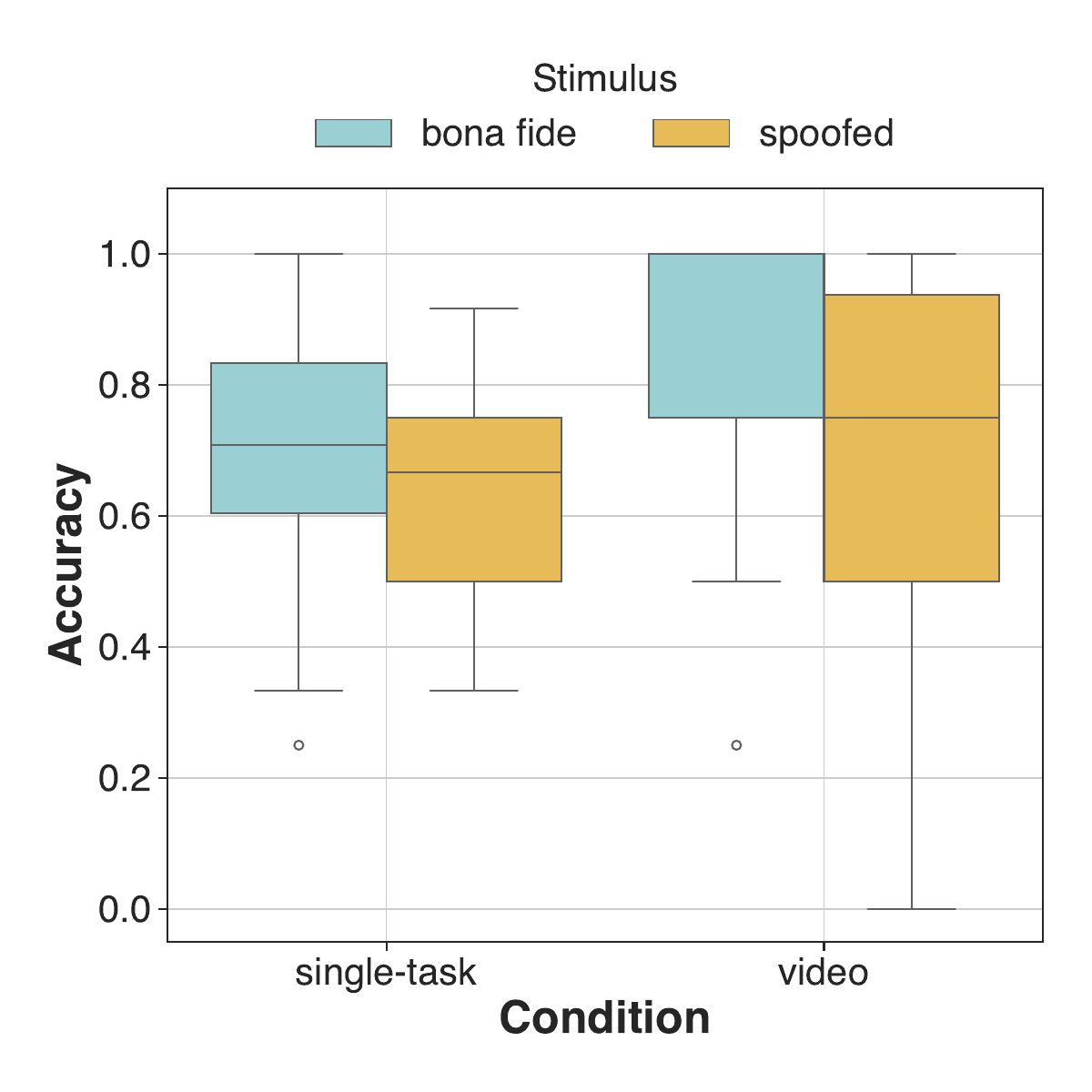}
\caption{Accuracy in detecting voice clones in the video condition of Experiment~2 in comparison to the single-task condition of Experiment~1 for spoofed and bona fide stimuli averaged by participant.}
\label{accuracy-per-condition-video}
\Description{The figure shows a box plot comparing accuracies in detecting voice clones between the single-task and video-task conditions. Furthermore, the detection accuracies are grouped by bona fide and spoofed stimuli. The medians for the single-task conditions are about 0.7 for bona fide and 0.66 for spoofed stimuli. Under the video condition, detection accuracy medians are 1.0 for bona fide and 0.78 for spoofed stimuli.}
\end{figure}

Participants in Experiment~2, in which a video is shown in parallel with bona fide and spoofed audio clips, achieved an average accuracy of 0.75. This accuracy is significantly higher than under the single ($\mu=0.68$, $p=0.03$) and dual-task ($\mu=0.66$, $p=0.005$) conditions in Experiment~1. In Figure~\ref{accuracy-per-condition-video}, the accuracies of detecting bona fide and spoofed stimuli in the video condition are compared. On average, participants detect bona fide stimuli with an accuracy of 0.78 and spoofed stimuli with an accuracy of 0.71. According to a t-test, these differences are not significant ($p=0.24$). Noteworthy are also the even larger standard deviations of 0.22 and 0.25 for bona fide and spoofed stimuli, respectively. 

As Experiment~2 was conducted after Experiment~1, we analyze if learning effects could explain the higher achieved accuracies. Independent of what the condition was (order was randomized), the average accuracy under the first condition is 0.65 while under the second condition is 0.68. Although higher accuracies are achieved in the second condition, the difference is not significant according to a paired t-test ($p=0.2$).

\begin{figure*}[ht]
\centering
\includegraphics[width=.9\linewidth,trim={0 40pt 0 20pt}]{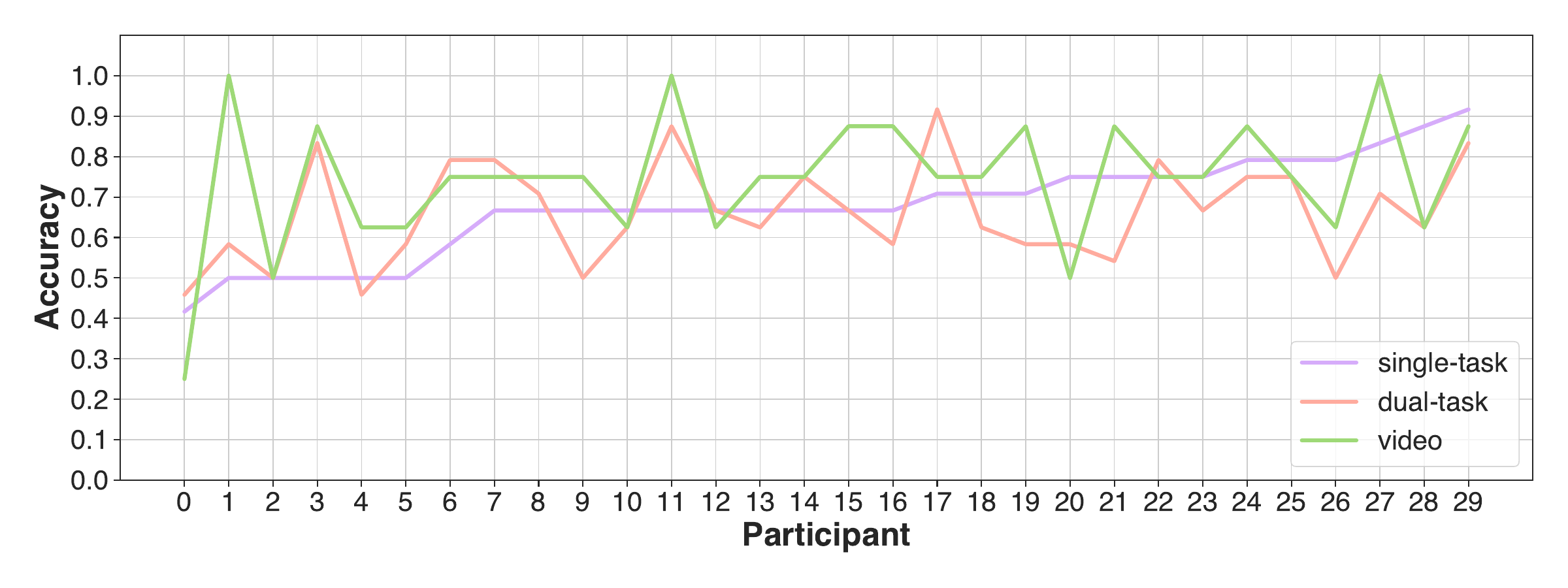}
\caption{Accuracy in detecting voice clones for each participant in the single-task, dual-task and video conditions. Participants sorted by single-task accuracy.}
\label{accuracy-per-participant}
\Description{The figure shows a line plot that visualizes individual participant's accuracies in the single-task, dual-task, and video conditions.}
\end{figure*}

Finally, we wondered whether the participants who perform well in Experiment~1 also perform well in Experiment~2. Figure~\ref{accuracy-per-participant} shows individual task performances of each participant in single-task, dual-task and video conditions. In the plot, we can observe that in many cases, participants perform similarly good or bad across Experiment~1 and 2 with a few exceptions. Participant 1 got a perfect accuracy in the video condition while only achieving 50\% accuracy in the single-task condition. Similarly, participant 3 did well in the dual-task and video conditions, but also only made 50\% correct guesses in the single-task condition. A correlation analysis reveals that there is a moderate Pearson correlation between accuracies of single-task versus dual-task ($r=0.32$), single-task versus video-task ($r=0.33$), and dual-task versus video-task ($r=0.49$).

\paragraph{Decision Indicators}

Although participants are shown visual stimuli at the same time, the decision indicators on which participants base their decisions do not differ substantially between Experiment~1 and Experiment~2. The only noticeable difference is that at least one participant stated that the videos made the stimulus more credible and prompted him to vote for ``bona fide.''

\section{Discussion}

Based on the results of our two conducted experiments, the impact of cognitive load on human detection abilities of voice-based deepfakes remains ambiguous. There is no significant difference in average accuracy achieved under single-task and dual-task conditions. However, we observe a drastic increase in standard deviation of the accuracy going from the single-task to the dual-task condition for detecting spoofed stimuli. We discuss three plausible theories which could explain these observations. 

The first theory entails that a 1-back task does not cause sufficient cognitive load to significantly impact the detection accuracy. This theory is backed up by the difficulty ratings of the participants, of which only 11 found the single-task easier than the dual-task. 13 participant found dual-task to be as easy as the single-task condition, while 6 participants even found the dual-task to be the easier condition. An argument against this theory is the found correlation between the primary and secondary task performance. Although this correlation is weak, it shows that some cognitive capacities have to be divided across both tasks. 

Given the fact that some participants found the dual-task condition easier than the single-task condition, our second theory states that a second stimulus or task helps some participants to focus more on the primary stimulus. For example, some studies found that background music can increase focus on sustained attention tasks \cite{kiss:2021}. Another example is the accessory stimulus effect, in which task-irrelevant ``accessory'' stimuli can reduce the reaction time of participants in a primary reaction task \cite{hackley:1999}. This theory would explain the significantly higher accuracy in the video condition in Experiment 2. On the contrary, the theory does not explain the variable accuracy of participants under the dual-task condition.

Assuming that the accessory stimulus effect applies for the video condition of Experiment~2 but not for the dual-task condition of Experiment~1, our third theory tries to explain the found accuracies of the dual-task condition. This theory hypothesizes that there is no interference between cognitive load caused by a 1-back task and voice-based deepfake detection. While the 1-back task makes use of working memory, detecting voice clones is an auditory processing task which might require a different kind of attention. If this theory is true, then the non-significant differences between single-task and dual-task condition can be considered random noise.  

The detection abilities of voice clones vary a lot from human to human, which gets evident by the overall high standard deviation across all tasks. The worst participant achieved an accuracy below chance in all three conditions, while the best participant maintained an accuracy above 80\% across all tasks. However, participants are somewhat consistent in their performance independent of the task as shown by their moderately correlating task accuracies. This shows how subjective the detection abilities are and that the different accuracies were not achieved due to draws of easy stimuli.

\section{Conclusion}

People often encounter deepfakes in situations in which they are exposed to cognitive load. Therefore, we examined the effect of cognitive load on the accuracy of detecting voice-based deepfakes with an empirical study. In two experiments, we compared the baseline human detection performance to the accuracy while solving a 1-back task in parallel or while watching symbolic video footage. We found that a light cognitive load as caused by a 1-back task does not systematically affect human detection accuracy. Furthermore, we observed that participants who watched video footage in parallel performed significantly better in the detection task, which could be related to the accessory stimuli effect. 

As the 1-back task caused potentially light cognitive load in the participants, we want to repeat the experiment with considerably harder secondary tasks. To determine whether these loads are realistic compared to social media use, we plan to add a social media simulation with scrolling through other posts and advertisements as an additional experiment. In our experiments, only a single stimulus-generating pipeline has been used in, which can be added as an independent variable in a follow-up study. Moreover, the role of the familiarity with the cloned speaker has not been analyzed yet, which could be subject of a future study. 

As we have observed, deepfakes can be hard for some people to identify. We encountered participants with below chance accuracies on the task. Supporting those people by educating them or implementing intervention strategies on social media platform is detrimental to protect our society from waves of disinformation. We hope that this and earlier studies are a call to action to mitigate the negative effects of disinformation and make social media platforms a safer place for information access.

\begin{acks}

This work was partially supported by the Federal Ministry of Research, Technology and Space (BMFTR) under grant 16KIS2470 (PADSE: Person-centered, audio- and voice-based deepfake and shallowfake detection).

\end{acks}

\bibliography{chiir26-lit}

@misc{abu-el-haija:2016,
  title = {{{YouTube-8M}}: {{A Large-Scale Video Classification Benchmark}}},
  author = {{Abu-El-Haija}, Sami and Kothari, Nisarg and Lee, Joonseok and Natsev, Paul and Toderici, George and Varadarajan, Balakrishnan and Vijayanarasimhan, Sudheendra},
  year = 2016,
  doi = {10.48550/arXiv.1609.08675}
}

@article{al-khazraji:2023,
  title = {Impact of {{Deepfake Technology}} on {{Social Media}}: {{Detection}}, {{Misinformation}} and {{Societal Implications}}},
  author = {{Al-khazraji}, Samer and Saleh, Hassan and Khalid, Adil and Mishkhal, Israa},
  year = 2023,
  month = oct,
  journal = {The Eurasia Proceedings of Science Technology Engineering and Mathematics},
  volume = {23},
  pages = {429--441},
  doi = {10.55549/epstem.1371792}
}

@misc{alali:2024,
  title = {An {{RFP}} Dataset for {{Real}}, {{Fake}}, and {{Partially}} Fake Audio Detection},
  author = {Alali, Abdulazeez and Theodorakopoulos, George},
  year = 2024,
  doi = {10.48550/arxiv.2404.17721}
}

@article{alali:2025,
  title = {Partial {{Fake Speech Attacks}} in the {{Real World Using Deepfake Audio}}},
  author = {Alali, Abdulazeez and Theodorakopoulos, George},
  year = 2025,
  journal = {Journal of Cybersecurity and Privacy},
  volume = {5},
  number = {1},
  numpages = {6},
  doi = {10.3390/JCP5010006}
}

@inproceedings{amirkhani:2025a,
  title = {Detecting the {{Undetectable}}: {{Human Judgments}} and the {{Challenge}} of {{Synthetic Voices}}},
  shorttitle = {Detecting the {{Undetectable}}},
  booktitle = {Proceedings of the 12th {{International Conference}} on {{Communities}} \& {{Technologies}} ({{C}}\&{{T}} 2025)},
  author = {Amirkhani, Sima and Stevens, Gunnar and Shajalal, M. D. and Boden, Alexander},
  year = 2025,
  publisher = {European Society for Socially Embedded Technologies (EUSSET)},
  address = {Siegen, Germany},
  doi = {10.48340/ct2025-1030},
  numpages = {6}
}

@inproceedings{bain:2023,
  title = {{{WhisperX}}: {{Time-Accurate Speech Transcription}} of {{Long-Form Audio}}},
  booktitle = {24th {{Annual Conference}} of the {{International Speech Communication Association}}, {{Interspeech}} 2023},
  author = {Bain, Max and Huh, Jaesung and Han, Tengda and Zisserman, Andrew},
  editor = {Harte, Naomi and {Carson-Berndsen}, Julie and Jones, Gareth},
  year = 2023,
  pages = {4489--4493},
  publisher = {ISCA},
  address = {Dublin, Ireland},
  doi = {10.21437/INTERSPEECH.2023-78}
}

@inproceedings{barnekow:2021,
  title = {Creation and {{Detection}} of {{German Voice Deepfakes}}},
  booktitle = {Foundations and {{Practice}} of {{Security}} - 14th {{International Symposium}}, {{FPS}} 2021},
  author = {Barnekow, Vanessa and Binder, Dominik and Kromrey, Niclas and Munaretto, Pascal and Schaad, Andreas and Schmieder, Felix},
  editor = {A{\"\i}meur, Esma and Laurent, Maryline and Yaich, Reda and Dupont, Beno{\^\i}t and {Garc{\'\i}a-Alfaro}, Joaqu{\'\i}n},
  year = 2021,
  series = {Lecture {{Notes}} in {{Computer Science}}},
  volume = {13291},
  pages = {355--364},
  publisher = {Springer},
  address = {Paris, France},
  doi = {10.1007/978-3-031-08147-7\_24}
}

@misc{barrington:2024,
  title = {{{DeepSpeak Dataset}} v1.0},
  author = {Barrington, Sarah and Bohacek, Matyas and Farid, Hany},
  year = 2024,
  doi = {10.48550/ARXIV.2408.05366}
}

@article{barrington:2025,
  title = {People Are Poorly Equipped to Detect {{AI-powered}} Voice Clones},
  author = {Barrington, Sarah and Cooper, Emily A. and Farid, Hany},
  year = 2025,
  month = mar,
  journal = {Scientific Reports},
  volume = {15},
  number = {1},
  pages = {11004},
  publisher = {Nature Publishing Group},
  issn = {2045-2322},
  doi = {10.1038/s41598-025-94170-3},
  copyright = {2025 The Author(s)}
}

@misc{bastian:2023,
  title = {{{AI-generated}} Fake Audio of {{Germany}}'s Top News Program "{{Tagesschau}}" Spreads Disinformation},
  author = {Bastian, Matthias},
  year = 2023,
  month = nov,
  journal = {The Decoder},
  howpublished = {https://the-decoder.com/ai-generated-fake-audio-of-germanys-top-news-program-tagesschau-spreads-disinformation/}
}

@book{bateman:2020,
  title = {Deepfakes and Synthetic Media in the Financial System: {{Assessing}} Threat Scenarios},
  author = {Bateman, Jon},
  year = 2020,
  month = jul,
  publisher = {Carnegie Endowment for International Peace},
  address = {Washington D.C., United States}
}

@misc{betker:2023,
  title = {Better Speech Synthesis through Scaling},
  author = {Betker, James},
  year = 2023,
  doi = {10.48550/arxiv.2305.07243},
}

@misc{bhalli:2024,
  title = {Listening for {{Expert Identified Linguistic Features}}: {{Assessment}} of {{Audio Deepfake Discernment}} among {{Undergraduate Students}}},
  author = {Bhalli, Noshaba Nasir and Naqvi, Nehal and Evered, Chloe and Mallinson, Christine and Janeja, Vandana P.},
  year = 2024,
  doi = {10.48550/arxiv.2411.14586},
}

@inproceedings{chen:2025,
  title = {F5-{{TTS}}: {{A Fairytaler}} That {{Fakes Fluent}} and {{Faithful Speech}} with {{Flow Matching}}},
  booktitle = {Proceedings of the 63rd {{Annual Meeting}} of the {{Association}} for {{Computational Linguistics}} ({{Volume}} 1: {{Long Papers}}), {{ACL}} 2025},
  author = {Chen, Yushen and Niu, Zhikang and Ma, Ziyang and Deng, Keqi and Wang, Chunhui and Zhao, Jian and Yu, Kai and Chen, Xie},
  editor = {Che, Wanxiang and Nabende, Joyce and Shutova, Ekaterina and Pilehvar, Mohammad Taher},
  year = 2025,
  pages = {6255--6271},
  publisher = {Association for Computational Linguistics},
  address = {Vienna, Austria}
}

@inproceedings{chong:2023,
  title = {Bot or {{Human}}? {{Detection}} of {{DeepFake Text}} with {{Semantic}}, {{Emoji}}, {{Sentiment}} and {{Linguistic Features}}},
  booktitle = {13th {{IEEE International Conference}} on {{System Engineering}} and {{Technology}}, {{ICSET}} 2023},
  author = {Chong, Alicia Tsui Ying and Chua, Hui Na and Jasser, Muhammed Basheer and Wong, Richard T. K.},
  year = 2023,
  pages = {205--210},
  publisher = {IEEE},
  address = {Shah Alam, Malaysia},
  doi = {10.1109/ICSET59111.2023.10295100}
}

@misc{cooke:2024,
  title = {As {{Good As A Coin Toss}}: {{Human}} Detection of {{AI-generated}} Images, Videos, Audio, and Audiovisual Stimuli},
  author = {Cooke, Di and Edwards, Abigail and Barkoff, Sophia and Kelly, Kathryn},
  year = 2024,
  doi = {10.48550/arxiv.2403.16760},
}

@techreport{deeptrace:2019,
  title = {The {{State}} of {{Deepfakes}}},
  author = {{Deeptrace}},
  year = 2019,
  month = sep
}

@inproceedings{demirsahin:2020,
  title = {Open-Source {{Multi-speaker Corpora}} of the {{English Accents}} in the {{British Isles}}},
  booktitle = {Proceedings of {{The}} 12th {{Language Resources}} and {{Evaluation Conference}}},
  author = {Demirsahin, Isin and Kjartansson, Oddur and Gutkin, Alexander and Rivera, Clara},
  editor = {Calzolari, Nicoletta and B{\'e}chet, Fr{\'e}d{\'e}ric and Blache, Philippe and Choukri, Khalid and Cieri, Christopher and Declerck, Thierry and Goggi, Sara and Isahara, Hitoshi and Maegaard, Bente and Mariani, Joseph and Mazo, H{\'e}l{\`e}ne and Moreno, Asunci{\'o}n and Odijk, Jan and Piperidis, Stelios},
  year = 2020,
  month = may,
  pages = {6532--6541},
  publisher = {European Language Resources Association},
  address = {Marseille, France},
  isbn = {979-10-95546-34-4},
}

@article{de-rancourt-raymond:2022,
  title = {The Unethical Use of Deepfakes},
  author = {{de Rancourt-Raymond}, Audrey and Smaili, Nadia},
  year = 2022,
  month = may,
  journal = {Journal of Financial Crime},
  volume = {30},
  number = {4},
  pages = {1066--1077},
  issn = {1359-0790},
  doi = {10.1108/JFC-04-2022-0090}
}

@article{findlay:2025,
  title = {`{{You}}'re Gonna Find This Creepy': My {{AI-cloned}} Voice Was Used by the Far Right. {{Could I}} Stop It?},
  author = {Findlay, Georgina},
  year = 2025,
  journal = {The Guardian},
  howpublished = {https://www.theguardian.com/commentisfree/2025/jan/07/ai-clone-voice-far-right-fake-audio},
}

@inproceedings{frank:2021,
  title = {{{WaveFake}}: {{A Data Set}} to {{Facilitate Audio Deepfake Detection}}},
  booktitle = {Proceedings of the {{Neural Information Processing Systems Track}} on {{Datasets}} and {{Benchmarks}} 1, {{NeurIPS Datasets}} and {{Benchmarks}} 2021},
  author = {Frank, Joel and Sch{\"o}nherr, Lea},
  editor = {Vanschoren, Joaquin and Yeung, Sai-Kit},
  year = 2021,
  publisher = {NeurIPS},
  volume = {1},
  address = {Virtual Event},
  numpages = {17}
}

@inproceedings{frank:2024,
  title = {A {{Representative Study}} on {{Human Detection}} of {{Artificially Generated Media Across Countries}}},
  booktitle = {{{IEEE Symposium}} on {{Security}} and {{Privacy}}, {{SP}} 2024},
  author = {Frank, Joel and Herbert, Franziska and Ricker, Jonas and Sch{\"o}nherr, Lea and Eisenhofer, Thorsten and Fischer, Asja and D{\"u}rmuth, Markus and Holz, Thorsten},
  year = 2024,
  pages = {55--73},
  publisher = {IEEE},
  address = {San Francisco, CA, USA},
  doi = {10.1109/SP54263.2024.00159}
}

@inproceedings{gohsen:2023,
  title = {Guiding {{Oral Conversations}}: {{How}} to {{Nudge Users Towards Asking Questions}}?},
  booktitle = {8th {{ACM SIGIR Conference}} on {{Human Information Interaction}} and {{Retrieval}} ({{CHIIR}} 2023)},
  author = {Gohsen, Marcel and Kiesel, Johannes and Korashi, Mariam and Ehlers, Jan and Stein, Benno},
  year = 2023,
  month = mar,
  pages = {34--42},
  publisher = {ACM},
  address = {New York, United States},
  doi = {10.1145/3576840.3578291},
  site = {Austin, TX, USA}
}

@inproceedings{gomez-rodriguez:2014,
  title = {Quantifying {{Information Overload}} in {{Social Media}} and {{Its Impact}} on {{Social Contagions}}},
  booktitle = {Proceedings of the {{Eighth International Conference}} on {{Weblogs}} and {{Social Media}}, {{ICWSM}} 2014},
  author = {{Gomez-Rodriguez}, Manuel and Gummadi, Krishna P. and Sch{\"o}lkopf, Bernhard},
  editor = {Adar, Eytan and Resnick, Paul and Choudhury, Munmun De and Hogan, Bernie and Oh, Alice},
  year = 2014,
  pages = {170--179},
  publisher = {The AAAI Press},
  address = {Ann Arbor, Michigan, USA},
  doi = {10.1609/icwsm.v8i1.14549 },
}

@inproceedings{grimmer:2021,
  title = {The {{Cognitive Eye}}: {{Indexing Oculomotor Functions}} for {{Mental Workload Assessment}} in {{Cognition-Aware Systems}}},
  booktitle = {{{CHI}} '21: {{CHI Conference}} on {{Human Factors}} in {{Computing Systems}}},
  author = {Grimmer, Janine and Simon, Laura and Ehlers, Jan},
  editor = {Kitamura, Yoshifumi and Quigley, Aaron and Isbister, Katherine and Igarashi, Takeo},
  year = 2021,
  pages = {428:1--428:6},
  publisher = {ACM},
  address = {Yokohama, Japan},
  doi = {10.1145/3411763.3451662}
}

@article{groh:2022,
  title = {Deepfake Detection by Human Crowds, Machines, and Machine-Informed Crowds},
  author = {Groh, Matthew and Epstein, Ziv and Firestone, Chaz and Picard, Rosalind},
  year = 2022,
  month = jan,
  journal = {Proceedings of the National Academy of Sciences},
  volume = {119},
  number = {1},
  pages = {e2110013119},
  publisher = {Proceedings of the National Academy of Sciences},
  doi = {10.1073/pnas.2110013119}
}

@article{groh:2024,
  title = {Human Detection of Political Speech Deepfakes across Transcripts, Audio, and Video},
  author = {Groh, Matthew and Sankaranarayanan, Aruna and Singh, Nikhil and Kim, Dong Young and Lippman, Andrew and Picard, Rosalind},
  year = 2024,
  month = sep,
  journal = {Nature Communications},
  volume = {15},
  number = {1},
  pages = {7629},
  publisher = {Nature Publishing Group},
  issn = {2041-1723},
  doi = {10.1038/s41467-024-51998-z},
  copyright = {2024 The Author(s)}
}

@article{hackley:1999,
  title = {Accessory {{Stimulus Effects}} on {{Response Selection}}: {{Does Arousal Speed Decision Making}}?},
  shorttitle = {Accessory {{Stimulus Effects}} on {{Response Selection}}},
  author = {Hackley, Steven A. and {Valle-Incl{\'a}n}, Fernando},
  year = 1999,
  month = may,
  journal = {Journal of Cognitive Neuroscience},
  volume = {11},
  number = {3},
  pages = {321--329},
  issn = {0898-929X},
  doi = {10.1162/089892999563427}
}

@inproceedings{han:2024,
  title = {Uncovering {{Human Traits}} in {{Determining Real}} and {{Spoofed Audio}}: {{Insights}} from {{Blind}} and {{Sighted Individuals}}},
  booktitle = {Proceedings of the {{CHI Conference}} on {{Human Factors}} in {{Computing Systems}}, {{CHI}} 2024},
  author = {Han, Chaeeun and Mitra, Prasenjit and Billah, Syed Masum},
  editor = {Mueller, Florian 'Floyd' and Kyburz, Penny and Williamson, Julie R. and Sas, Corina and Wilson, Max L. and Dugas, Phoebe O. Toups and Shklovski, Irina},
  year = 2024,
  pages = {949:1--949:14},
  publisher = {ACM},
  address = {Honolulu, HI, USA},
  doi = {10.1145/3613904.3642817}
}

@misc{hashmi:2024,
  title = {Unmasking {{Illusions}}: {{Understanding Human Perception}} of {{Audiovisual Deepfakes}}},
  author = {Hashmi, Ammarah and Shahzad, Sahibzada Adil and Lin, Chia-Wen and Tsao, Yu and Wang, Hsin-Min},
  year = 2024,
  doi = {10.48550/arxiv.2405.04097}
}

@inproceedings{hodas:2012,
  title = {How {{Visibility}} and {{Divided Attention Constrain Social Contagion}}},
  booktitle = {2012 {{International Conference}} on {{Privacy}}, {{Security}}, {{Risk}} and {{Trust}}, {{PASSAT}} 2012, and 2012 {{International Confernece}} on {{Social Computing}}, {{SocialCom}} 2012},
  author = {Hodas, Nathan Oken and Lerman, Kristina},
  year = 2012,
  pages = {249--257},
  publisher = {IEEE Computer Society},
  address = {Amsterdam, Netherlands},
  doi = {10.1109/SOCIALCOM-PASSAT.2012.129}
}

@inproceedings{huber:2019,
  title = {B-{{Script}}: {{Transcript-based B-roll Video Editing}} with {{Recommendations}}},
  booktitle = {Proceedings of the 2019 {{CHI Conference}} on {{Human Factors}} in {{Computing Systems}}, {{CHI}} 2019},
  author = {Huber, Bernd and Shin, Hijung Valentina and Russell, Bryan C. and Wang, Oliver and Mysore, Gautham J.},
  editor = {Brewster, Stephen A. and Fitzpatrick, Geraldine and Cox, Anna L. and Kostakos, Vassilis},
  year = 2019,
  pages = {81},
  publisher = {ACM},
  address = {Glasgow, Scotland, UK},
  doi = {10.1145/3290605.3300311}
}

@misc{ito:2017,
  title = {The {{LJ Speech Dataset}}},
  author = {Ito, Keith and Johnson, Linda},
  year = 2017,
  howpublished = {https://keithito.com/LJ-Speech-Dataset/},
}

@misc{josephs:2023,
  title = {Artifact Magnification on Deepfake Videos Increases Human Detection and Subjective Confidence},
  author = {Josephs, Emilie and Fosco, Camilo and Oliva, Aude},
  year = 2023,
  doi = {10.48550/arxiv.2304.04733}
}

@inproceedings{khalid:2021,
  title = {{{FakeAVCeleb}}: {{A Novel Audio-Video Multimodal Deepfake Dataset}}},
  booktitle = {Proceedings of the {{Neural Information Processing Systems Track}} on {{Datasets}} and {{Benchmarks}} 1, {{NeurIPS Datasets}} and {{Benchmarks}} 2021},
  author = {Khalid, Hasam and Tariq, Shahroz and Kim, Minha and Woo, Simon S.},
  editor = {Vanschoren, Joaquin and Yeung, Sai-Kit},
  year = 2021,
  publisher = {NeurIPS},
  address = {Virtual Event},
  volume = {1},
  numpages = {14}
}

@misc{kim:2021a,
  title = {{{KUIELab-MDX-Net}}: {{A Two-Stream Neural Network}} for {{Music Demixing}}},
  author = {Kim, Minseok and Choi, Woo-Sung and Chung, Jaehwa and Lee, Daewon and Jung, Soonyoung},
  year = 2021,
  doi = {10.48550/arXiv.2111.12203},
}

@inproceedings{kinnunen:2017,
  title = {The {{ASVspoof}} 2017 {{Challenge}}: {{Assessing}} the {{Limits}} of {{Replay Spoofing Attack Detection}}},
  booktitle = {18th {{Annual Conference}} of the {{International Speech Communication Association}}, {{Interspeech}} 2017},
  author = {Kinnunen, Tomi and Sahidullah, {\relax Md}. and Delgado, H{\'e}ctor and Todisco, Massimiliano and Evans, Nicholas W. D. and Yamagishi, Junichi and Lee, Kong-Aik},
  editor = {Lacerda, Francisco},
  year = 2017,
  pages = {2--6},
  publisher = {ISCA},
  address = {Stockholm, Sweden},
  doi = {10.21437/INTERSPEECH.2017-1111}
}

@article{kiss:2021,
  title = {The Effect of Preferred Background Music on Task-Focus in Sustained Attention},
  author = {Kiss, Luca and Linnell, Karina J.},
  year = 2021,
  month = sep,
  journal = {Psychological Research},
  volume = {85},
  number = {6},
  pages = {2313--2325},
  issn = {1430-2772},
  doi = {10.1007/s00426-020-01400-6}
}

@misc{kominek:2003,
  title = {{{CMU ARCTIC}} Databases for Speech Synthesis},
  author = {Kominek, John and Black, Alan W.},
  year = 2003,
  address = {Carnegie Mellon University},
  howpublished = {http://www.festvox.org/cmu\_arctic/},
}

@article{lazer:2018,
  title = {The Science of Fake News},
  author = {Lazer, David M. J. and Baum, Matthew A. and Benkler, Yochai and Berinsky, Adam J. and Greenhill, Kelly M. and Menczer, Filippo and Metzger, Miriam J. and Nyhan, Brendan and Pennycook, Gordon and Rothschild, David and Schudson, Michael and Sloman, Steven A. and Sunstein, Cass R. and Thorson, Emily A. and Watts, Duncan J. and Zittrain, Jonathan L.},
  year = 2018,
  month = mar,
  journal = {Science},
  volume = {359},
  number = {6380},
  pages = {1094--1096},
  publisher = {American Association for the Advancement of Science},
  doi = {10.1126/science.aao2998}
}

@inproceedings{li:2023a,
  title = {{{StyleTTS}} 2: {{Towards Human-Level Text-to-Speech}} through {{Style Diffusion}} and {{Adversarial Training}} with {{Large Speech Language Models}}},
  booktitle = {Advances in {{Neural Information Processing Systems}} 36: {{Annual Conference}} on {{Neural Information Processing Systems}} 2023, {{NeurIPS}} 2023},
  author = {Li, Yinghao Aaron and Han, Cong and Raghavan, Vinay S. and Mischler, Gavin and Mesgarani, Nima},
  editor = {Oh, Alice and Naumann, Tristan and Globerson, Amir and Saenko, Kate and Hardt, Moritz and Levine, Sergey},
  year = 2023,
  publisher = {Curran Associates, Inc.},
  address = {New Orleans, LA, USA},
  volume = {36},
  pages = {19594--19621},
}

@article{liu:2023a,
  title = {{{ASVspoof}} 2021: {{Towards Spoofed}} and {{Deepfake Speech Detection}} in the {{Wild}}},
  author = {Liu, Xuechen and Wang, Xin and Sahidullah, {\relax Md}. and Patino, Jose and Delgado, H{\'e}ctor and Kinnunen, Tomi and Todisco, Massimiliano and Yamagishi, Junichi and Evans, Nicholas W. D. and Nautsch, Andreas and Lee, Kong Aik},
  year = 2023,
  journal = {IEEE ACM Trans. Audio Speech Lang. Process.},
  volume = {31},
  pages = {2507--2522},
  doi = {10.1109/TASLP.2023.3285283}
}

@inproceedings{lopez-garcia:2024,
  title = {Speaking with {{Objects}}: {{Conversational Agents}}' {{Embodiment}} in {{Virtual Museums}}},
  booktitle = {2024 {{IEEE International Symposium}} on {{Mixed}} and {{Augmented Reality}} ({{ISMAR}} 2024)},
  author = {L{\'o}pez Garc{\'i}a, Irene and Schott, Ephraim and Gohsen, Marcel and Bernhard, Volker and Stein, Benno and Fr{\"o}hlich, Bernd},
  editor = {Eck, Ulrich and Sra, Misha and Stefanucci, Jeanine K. and Sugimoto, Maki and Tatzgern, Markus and Williams, Ian},
  year = 2024,
  month = oct,
  pages = {279--288},
  publisher = {IEEE},
  address = {Greater Seattle Area, USA},
  doi = {10.1109/ISMAR62088.2024.00042},
  site = {Greater Seattle Area, USA}
}

@misc{maclean:2018,
  title = {Voxforge},
  author = {MacLean, Ken},
  year = 2018,
  howpublished = {https://www.voxforge.org/home}
}

@article{mai:2023,
  title = {Warning: {{Humans}} Cannot Reliably Detect Speech Deepfakes},
  shorttitle = {Warning},
  author = {Mai, Kimberly T. and Bray, Sergi and Davies, Toby and Griffin, Lewis D.},
  year = 2023,
  month = aug,
  journal = {PLOS ONE},
  volume = {18},
  number = {8},
  pages = {e0285333},
  publisher = {Public Library of Science},
  issn = {1932-6203},
  doi = {10.1371/journal.pone.0285333}
}

@article{mania:2024,
  title = {Legal {{Protection}} of {{Revenge}} and {{Deepfake Porn Victims}} in the {{European Union}}: {{Findings From}} a {{Comparative Legal Study}}},
  shorttitle = {Legal {{Protection}} of {{Revenge}} and {{Deepfake Porn Victims}} in the {{European Union}}},
  author = {Mania, Karolina},
  year = 2024,
  month = jan,
  journal = {Trauma, Violence, \& Abuse},
  volume = {25},
  number = {1},
  pages = {117--129},
  publisher = {SAGE Publications},
  issn = {1524-8380},
  doi = {10.1177/15248380221143772}
}

@article{mirsky:2022,
  title = {The {{Creation}} and {{Detection}} of {{Deepfakes}}: {{A Survey}}},
  author = {Mirsky, Yisroel and Lee, Wenke},
  year = 2022,
  journal = {Acm Computing Surveys},
  volume = {54},
  number = {1},
  pages = {7:1--7:41},
  doi = {10.1145/3425780}
}

@inproceedings{muller:2022,
  title = {Human {{Perception}} of {{Audio Deepfakes}}},
  booktitle = {{{DDAM}}@{{MM}} 2022: {{Proceedings}} of the 1st {{International Workshop}} on {{Deepfake Detection}} for {{Audio Multimedia}}},
  author = {M{\"u}ller, Nicolas M. and Pizzi, Karla and Williams, Jennifer},
  editor = {Tao, Jianhua and Li, Haizhou and Meng, Helen and Yu, Dong and Akagi, Masato and Yi, Jiangyan and Fan, Cunhang and Fu, Ruibo and Lian, Shan and Zhang, Pengyuan},
  year = 2022,
  pages = {85--91},
  publisher = {ACM},
  address = {Lisboa, Portugal},
  doi = {10.1145/3552466.3556531}
}

@inproceedings{peng:2024,
  title = {{{VoiceCraft}}: {{Zero-Shot Speech Editing}} and {{Text-to-Speech}} in the {{Wild}}},
  booktitle = {Proceedings of the 62nd {{Annual Meeting}} of the {{Association}} for {{Computational Linguistics}} ({{Volume}} 1: {{Long Papers}}), {{ACL}} 2024},
  author = {Peng, Puyuan and Huang, Po-Yao and Li, Shang-Wen and Mohamed, Abdelrahman and Harwath, David},
  editor = {Ku, Lun-Wei and Martins, Andre and Srikumar, Vivek},
  year = 2024,
  pages = {12442--12462},
  publisher = {Association for Computational Linguistics},
  address = {Bangkok, Thailand},
  doi = {10.18653/V1/2024.ACL-LONG.673}
}

@article{pittman:2023,
  title = {Cognitive {{Load}} and {{Social Media Advertising}}},
  author = {Pittman, Matthew and Haley, Eric},
  year = 2023,
  month = jan,
  journal = {Journal of Interactive Advertising},
  volume = {23},
  number = {1},
  pages = {33--54},
  publisher = {Routledge},
  issn = {null},
  doi = {10.1080/15252019.2022.2144780}
}

@inproceedings{prasad:2022,
  title = {Human vs. {{Automatic Detection}} of {{Deepfake Videos Over Noisy Channels}}},
  booktitle = {{{IEEE International Conference}} on {{Multimedia}} and {{Expo}}, {{ICME}} 2022},
  author = {Prasad, Swaroop Shankar and Hadar, Ofer and Vu, Thang and Polian, Ilia},
  year = 2022,
  pages = {1--6},
  publisher = {IEEE},
  address = {Taipei, Taiwan},
  doi = {10.1109/ICME52920.2022.9859954}
}

@inproceedings{reimao:2019,
  title = {{{FoR}}: {{A Dataset}} for {{Synthetic Speech Detection}}},
  booktitle = {2019 {{International Conference}} on {{Speech Technology}} and {{Human-Computer Dialogue}}, {{SpeD}} 2019},
  author = {Reimao, Ricardo and Tzerpos, Vassilios},
  editor = {Burileanu, Corneliu and Teodorescu, Horia-Nicolai},
  year = 2019,
  pages = {1--10},
  publisher = {IEEE},
  address = {Timisoara, Romania},
  doi = {10.1109/SPED.2019.8906599}
}

@inproceedings{sankaranarayanan:2021,
  address = {Aachen, Germany},
  title = {The {{Presidential Deepfakes Dataset}}},
  booktitle = {{{CEUR Workshop Proceedings}}},
  author = {Sankaranarayanan, Aruna and Groh, Matthew and Picard, Rosalind and Lippman, Andrew},
  year = 2021,
  volume = {2942},
  pages = {57--72},
  publisher = {CEUR-WS}
}

@inproceedings{segbroeck:2020,
  title = {{{DiPCo}} - {{Dinner Party Corpus}}},
  booktitle = {21st {{Annual Conference}} of the {{International Speech Communication Association}}, {{Interspeech}} 2020},
  author = {Segbroeck, Maarten Van and Zaid, Ahmed and Kutsenko, Ksenia and Huerta, Cirenia and Nguyen, Tinh and Luo, Xuewen and Hoffmeister, Bj{\"o}rn and Trmal, Jan and Omologo, Maurizio and Maas, Roland},
  editor = {Meng, Helen and Xu, Bo and Zheng, Thomas Fang},
  year = 2020,
  pages = {434--436},
  publisher = {ISCA},
  address = {Shanghai, China},
  doi = {10.21437/INTERSPEECH.2020-2800}
}

@inproceedings{sharevski:2024,
  title = {Blind and {{Low-Vision Individuals}}' {{Detection}} of {{Audio Deepfakes}}},
  booktitle = {Proceedings of the 2024 on {{ACM SIGSAC Conference}} on {{Computer}} and {{Communications Security}}, {{CCS}} 2024},
  author = {Sharevski, Filipo and Zeidieh, Aziz and Loop, Jennifer Vander and Jachim, Peter},
  editor = {Luo, Bo and Liao, Xiaojing and Xu, Jun and Kirda, Engin and Lie, David},
  year = 2024,
  pages = {4867--4881},
  publisher = {ACM},
  address = {Salt Lake City, UT, USA},
  doi = {10.1145/3658644.3690305}
}

@article{shu:2020,
  title = {Combating Disinformation in a Social Media Age},
  author = {Shu, Kai and Bhattacharjee, Amrita and Alatawi, Faisal and Nazer, Tahora H. and Ding, Kaize and Karami, Mansooreh and Liu, Huan},
  year = 2020,
  journal = {WIREs Data Mining and Knowledge Discovery},
  pages = {e1385},
  volume = {10},
  number = {6},
  doi = {10.1002/WIDM.1385}
}

@article{somoray:2023,
  title = {Providing Detection Strategies to Improve Human Detection of Deepfakes: {{An}} Experimental Study},
  author = {Somoray, Klaire and Miller, Dan J.},
  year = 2023,
  journal = {Computers in Human Behavior},
  volume = {149},
  pages = {107917},
  doi = {10.1016/J.CHB.2023.107917}
}

@article{wang:2020,
  title = {{{ASVspoof}} 2019: {{A}} Large-Scale Public Database of Synthesized, Converted and Replayed Speech},
  author = {Wang, Xin and Yamagishi, Junichi and Todisco, Massimiliano and Delgado, H{\'e}ctor and Nautsch, Andreas and Evans, Nicholas W. D. and Sahidullah, {\relax Md}. and Vestman, Ville and Kinnunen, Tomi and Lee, Kong Aik and Juvela, Lauri and Alku, Paavo and Peng, Yu-Huai and Hwang, Hsin-Te and Tsao, Yu and Wang, Hsin-Min and Maguer, S{\'e}bastien Le and Becker, Markus and Ling, Zhen-Hua},
  year = 2020,
  journal = {Computer Speech and Language},
  volume = {64},
  pages = {101114},
  doi = {10.1016/J.CSL.2020.101114}
}

@article{wang:2022a,
  title = {Users' Emotional and Behavioral Responses to Deepfake Videos of {{K-pop}} Idols},
  author = {Wang, Soyoung and Kim, Seongcheol},
  year = 2022,
  journal = {Computers in Human Behavior},
  volume = {134},
  pages = {107305},
  doi = {10.1016/J.CHB.2022.107305}
}

@inproceedings{warren:2024,
  title = {"{{Better Be Computer}} or {{I}}'m {{Dumb}}": {{A Large-Scale Evaluation}} of {{Humans}} as {{Audio Deepfake Detectors}}},
  booktitle = {Proceedings of the 2024 on {{ACM SIGSAC Conference}} on {{Computer}} and {{Communications Security}}, {{CCS}} 2024},
  author = {Warren, Kevin and Tucker, Tyler and Crowder, Anna and Olszewski, Daniel and Lu, Allison and Fedele, Caroline and Pasternak, Magdalena and Layton, Seth and Butler, Kevin R. B. and Gates, Carrie and Traynor, Patrick},
  editor = {Luo, Bo and Liao, Xiaojing and Xu, Jun and Kirda, Engin and Lie, David},
  year = 2024,
  pages = {2696--2710},
  publisher = {ACM},
  address = {Salt Lake City, UT, USA},
  doi = {10.1145/3658644.3670325}
}

@misc{watson:2021,
  title = {Audio {{Deepfake Perceptions}} in {{College Going Populations}}},
  author = {Watson, Gabrielle and Khanjani, Zahra and Janeja, Vandana P.},
  year = 2021,
  doi = {arXiv.2112.03351},
}

@inproceedings{wynn:2021,
  title = {Deepfake {{Portraits}} in {{Augmented Reality}} for {{Museum Exhibits}}},
  booktitle = {{{IEEE International Symposium}} on {{Mixed}} and {{Augmented Reality Adjunct}}, {{ISMAR}} 2021 {{Adjunct}}},
  author = {Wynn, Nathan and Johnsen, Kyle and Gonzalez, Nick},
  year = 2021,
  pages = {513--514},
  publisher = {IEEE},
  address = {Bari, Italy},
  doi = {10.1109/ISMAR-ADJUNCT54149.2021.00125}
}

@misc{yamagishi:2019,
  title = {{{CSTR VCTK Corpus}}: {{English Multi-speaker Corpus}} for {{CSTR Voice Cloning Toolkit}}},
  author = {Yamagishi, Junichi and Veaux, Christophe and MacDonald, Kirsten},
  year = 2019,
  month = nov,
  publisher = {University of Edinburgh. The Centre for Speech Technology Research (CSTR)},
  doi = {10.7488/ds/2645}
}

@inproceedings{zaramella:2023,
  title = {Why {{Don}}'t {{You Speak}}?: {{A Smartphone Application}} to {{Engage Museum Visitors Through Deepfakes Creation}}},
  booktitle = {Proceedings of the 5th {{Workshop}} on {{analySis}}, {{Understanding}} and {{proMotion}} of {{heritAge Contents}}, {{SUMAC}} 2023},
  author = {Zaramella, Matteo and Amerini, Irene and Russo, Paolo},
  editor = {{Gouet-Brunet}, Val{\'e}rie and Kosti, Ronak and Weng, Li},
  year = 2023,
  pages = {29--37},
  publisher = {ACM},
  address = {Ottawa, ON, Canada},
  doi = {10.1145/3607542.3617359}
}

@article{zhang:2020a,
  title = {An Overview of Online Fake News: {{Characterization}}, Detection, and Discussion},
  shorttitle = {An Overview of Online Fake News},
  author = {Zhang, Xichen and Ghorbani, Ali A.},
  year = 2020,
  month = mar,
  journal = {Information Processing \& Management},
  volume = {57},
  number = {2},
  pages = {102025},
  issn = {0306-4573},
  doi = {10.1016/j.ipm.2019.03.004}
}
\bibliographystyle{ACM-Reference-Format}

\end{document}